\documentclass[12pt, twoside]{article}
\newcommand{\bpart}{\mbox{\boldmath $\partial$}}
\newcommand{\bsigma}{\mbox{\boldmath $\sigma$}}
\newcommand{\balpha}{\mbox{\boldmath $\alpha$}}
\newcommand{\bgamma}{\mbox{\boldmath $\gamma$}}

\begin{document}

\title{\bf Fractionalization of angular momentum \\
at finite temperature around \\ a magnetic vortex}

\author{Yu.A. Sitenko and N.D. Vlasii}
\date{}
\maketitle
\begin{center}
{\small Bogolyubov Institute for Theoretical Physics,\\
National Academy of Sciences of Ukraine,\\
14-b Metrologichna str., Kyiv 03143, Ukraine\\
and\\
Physics Department, National Taras Shevchenko University of Kyiv,\\
2 Academician Glushkov ave., Kyiv 03127, Ukraine}
\end{center}

\begin{abstract}

Ambiguities in the definition of angular momentum of a
quantum-mechanical particle in the presence of a magnetic vortex are
reviewed. We show that the long-standing problem of the adequate
definition is resolved in the framework of the second-quantized
theory at nonzero temperature. Planar relativistic Fermi gas in the
background of a point-like magnetic vortex with arbitrary flux is
considered, and we find thermal averages, quadratic fluctuations,
and correlations of all observables, including angular momentum, in
this system. The kinetic definition of angular momentum is picked
out unambiguously by the requirement of plausible behaviour for the
angular momentum fluctuation and its correlation with fermion
number.

\end{abstract}

\begin{center}

fractional quantum numbers

\bigskip


\end{center}

\newpage
\section{Introduction}

Whereas in classical theory a comprehensive description of
electromagnetism is given in terms of the electromagnetic field
strength acting locally and directly on charged matter, this is not
the case in quantum theory. Since the quantum-mechanical wave
equation for a charged particle involves the electromagnetic vector
potential, the state of charged matter can be influenced by
electromagnetic effects even in the situation when the space-time
region of nonvanishing field strength is not accessible to charged
matter \cite{Ehre, Aha}. The latter was demonstrated for the case of
quantum-mechanical scattering on a tube of the magnetic force lines,
which was impenetrable for scattered particles: the differential
scattering cross section was found to be a periodic function of the
total magnetic flux confined in the tube \cite{Aha}. This phenomenon
with all its various generalizations (see, e.g., Refs.\cite{Pes,
SiM, Aud}) has no classical analogues and is denoted usually as the
Bohm-Aharonov effect.

Fractionalization of quantum numbers is a well-known feature of
various field theory models that describe interaction between
fermions and topological solitons \cite{Jac6, Su, Jac1, Gol, Nie3,
Jac4, Par}. It is usually convenient to interpret solitons as
background fields and to quantize fermion fields in such
backgrounds. In this respect it seems to be of interest to consider
a case when the region where a background field is nonvanishing does
not overlap with the region where a fermion field is quantized.
Namely, the quantized fermion matter does not penetrate a tube of
the magnetic force lines; for brevity, the impenetrable magnetic
tube will be called the magnetic vortex in the following. One
inquires about the dependence of the properties of this
second-quantized system on the vortex flux and the boundary
condition on the edge of the vortex; if a transverse size of the
vortex is neglected, then a parameter of the boundary condition
exhibits itself as a parameter of a self-adjoint extension of the
one-particle hamiltonian operator.

A study of this second-quantized system started two decades ago
\cite{Ser, Si8}. It was shown for a particular choice of the
boundary condition that fermion number \cite{Si0}, magnetic flux
\cite{Gor}, and angular momentum \cite{SiR} are induced in the
vacuum of quantized massive fermions. The induced vacuum quantum
numbers under the most general set of boundary conditions were
obtained in Refs.\cite{Si6, Si7, Sit9, Si9}. All results, except
angular momentum, are periodic in the value of the vortex flux, and
such a type of behaviour is consistent with the above mentioned for
the quantum-mechanical Bohm-Aharonov effect. As to angular momentum,
the situation is not so far determinative, which is due to two
alternatives for its definition. In quantum field theory, the
angular momentum operator, as well as operators of all other
observables, is obtained by sandwiching the appropriate
quantum-mechanical (one-particle) operator between the fermion field
operators (see Section 4 below). In quantum mechanics, there are two
possibilities to define angular momentum of a fermion particle in
the background of the Bohm--Aharonov magnetic vortex: the canonical
one that is quantized with half-integer eigenvalues,
\begin{equation}
\tilde{j}=n+\frac12,
\end{equation}
and the kinetic one that is quantized otherwise,
\begin{equation}
j=n+\frac{1}{2}-e\Phi,
\end{equation}
where $e$ is the electric charge and $\Phi$ is the vortex flux in
units of $2\pi$; strictly speaking, $\Phi$ is measured in units of
$2\pi\hbar c$, but we use conventional units $\hbar=c=1$. The
motivation in favour of the canonical definition lies in that it
yields angular momentum which is conserved when the vortex flux is
varied in time \cite{Pes, Gold, Lip, Jac3}. On the other hand, the
kinetic definition proves to be rather rewarding, since it leads to
such prolific concepts as anyons and fractional statistics
\cite{Wil2, Wilc} which are of fundamental value and possess
fascinating phenomenological applications to some condensed matter
systems \cite{Aro, Lau, Chen, Wil1}. It should be noted that
quantum-mechanical consequences of the kinetic definition are
periodic in the value of $e\Phi$ with the period equal to 1, and
this can be regarded as one more of manifestations of the
Bohm-Aharonov effect. If we turn to quantum field theory and
consider angular momentum which is induced in the vacuum around a
magnetic vortex, then the kinetic definition yields the quantity
which depends on the fractional part of $e\Phi$, while the canonical
definition yields the quantity which depends on both the fractional
and integer parts of $e\Phi$ \cite{SiR, Si9}.

In the present paper we shall study the effect of nonzero
temperature on the induced angular momentum around a magnetic
vortex, and consider its thermal average and quadratic fluctuation,
as well as correlations with other observables. The previous
analysis in Refs.\cite{Si4, Si5} is augmented by taking account of
the bulk volume contribution to thermal characteristics. In general,
the expressions for thermal characteristics consist of two pieces:
the one corresponding to the ideal gas contribution depends
essentially on a size of the system, increasing by power law as the
size increases, and the other one corresponding to the correction
due to interaction with a magnetic vortex is finite in the limit of
the infinite size; although, sometimes, as are the cases of the
averages of fermion number and induced magnetic flux and their
correlation, the ideal gas contribution is vanishing. We shall show
that the canonical definition of angular momentum yields rather
unconvincing results for the angular momentum fluctuation and its
correlation with fermion number: these quantities contain pieces
that are proportional to both powers of the system size and powers
of the whole vortex flux, and such a mismatch of different
contributions is hardly tolerable. On the contrary, the kinetic
definition of angular momentum yields plausible results, and this
allows one to conclude that quantum field theory at nonzero
temperature favours for certain the latter definition.

In the next section, we review the definition of angular momentum
for a quantum-mechanical charged particle placed into a classical
static rotationally invariant magnetic field. In Section 3 we
consider a situation when the particle does not penetrate the
magnetic field region and discuss some manifestations of
non-simply-connectedness of the accessible to the particle region. A
general formalism for the treatment of thermal characteristics of
observables in the framework of quantum field theory at finite
temperature is introduced in Section 4. Section 5 deals with planar
relativistic Fermi gas in the background of a magnetic vortex, and
we determine relevant one-particle spectral densities. Using
formalism of Section 4 and results of Section 5, we obtain thermal
characteristics of all observables in this system in Section 6. We
discuss our results in Section 7. Some details in the derivation of
the results are outlined in Appendices A-D.

\section{Angular momentum of a particle in a magnetic field}

Angular momentum of a free particle consists of the orbital and spin
parts
\begin{equation}
{\bf J}_0=-i\,{\bf x}\times{\bpart}+\frac{1}{2}{\bsigma}.
\end{equation}
If a charged particle is posited into an external electromagnetic
field, then a contribution of the latter has to be added to Eq.(3).
It can be shown (see Appendix A, Eq.(A.18)) that this contribution
in its turn consists of the purely field part
\begin{equation}
{\bf M}_{EM}=\int d^3x'\,({\bf x'}\times ({\bf E}\times {\bf B})),
\end{equation}
and the particle-field interaction part
\begin{equation}
{\bf M}_I=-\int d^3 x'\,{\bf x'}\times {\bf A}\,j^0=-e\,{\bf
x}\times {\bf A}({\bf x}),
\end{equation}
where
\begin{equation}
j^0=e\, \delta^3({\bf x'}-{\bf x})
\end{equation}
is the charge density of a particle located at point $\bf {x}$.
Adding Eq.(5) to Eq.(3) results in the emergence of the covariant
derivative, $\bpart\rightarrow\bpart-ie{\bf{A}}({\bf x})$, in the
orbital part.

Let us dwell now on the purely field contribution to the angular
momentum, Eq.(4). Substituting ${\bf B}={\bpart}\times{\bf A}_f$
(where ${\bf A}_f$ is a certain, fixed, vector potential) into
Eq.(4) and integrating it by parts, we get
\begin{eqnarray}
{M}^i_{EM}&\!\!\!\!=&\!\!\!\!\int
d^3x'\,\partial'_l(\varepsilon^{ikl}{x'}^k A^m_f
E^m-\varepsilon^{ikm}{x'}^k A^l_f E^m-\varepsilon^{ikm}{x'}^k
A^m_f E^l) + \nonumber
\\  &\!\!\!\!+& \!\!\!\!\int d^3x'\varepsilon^{ikm}{x'}^k[A^m_f\partial'_n
E^n+E^m\partial'_nA^n_f+A^n_f(\partial'_nE^m-\partial'_mE^n)].
\end{eqnarray}
Note that ${\bf M}_{EM}$ is the improved angular momentum which
differs from the canonical angular momentum ${\bf
\widetilde{M}}_{EM}$ (see Eqs.(A.11) and (A.12)),
\begin{equation}
M^i_{EM}-\widetilde{M}^i_{EM}=\int
d^3x'\,\partial'_l\varepsilon^{ikm}{x'}^k\chi^{m0l}=-\int
d^3x'\,\partial'_l\varepsilon^{ikm}{x'}^k A^mE^l,
\end{equation}
where we have used Eq.(A.16) for $\chi^{\mu\nu\lambda}$. We
consider a static rotationally invariant magnetic field, $({\bf
x}\times {\bpart})\cdot {\bf B}=0$, and choose ${\bf A}_f$
satisfying conditions
\begin{equation}
{\bf x}\cdot {\bf A}_f=0,\,\,\,{\bpart}\cdot{\bf A}_f=0.
\end{equation}
As to the electric field, it is generated by static charge density
$j^0$ (6),
\begin{equation}
{\bpart}'\cdot{\bf E}=j^0,\,\,\,\,\, {\bf
E}=\frac{e}{4\pi}\frac{{\bf x}'-{\bf x}}{|{\bf x}'-{\bf x}|^3},
\end{equation}
and is curlless, ${\bpart'}\times {\bf E}=0$. Then only the first
term in square brackets in the second integral in the right hand
side of Eq.(7) survives, yielding upon integration a contribution
which is equal to the contribution of particle-field interaction,
Eq.(5), taken with opposite sign and at ${\bf A}={\bf A}_f$. As to
the first integral in the right hand side of Eq.(7), it is
transformed to an integral over a closed surface, as described in
Appendix A, see Eqs.(A.4) and (A.13). Thus the behaviour of terms in
brackets in this integral is crucial: the first two terms decrease
as $|{\bf x'}|^{-3}$ at large $|{\bf x'}|$ yielding vanishing
contribution upon integration, whereas the last one decreases as
$|{\bf x'}|^{-2}$ at large $|{\bf x'}|$. This yields upon
integration finite contribution which is equal to the total magnetic
flux times $e(2\pi)^{-1}$ with the opposite sign \cite{Jac3}.
Consequently, one gets
\begin{equation}
{\bf M}_{EM}=e\,{\bf x}\times{\bf A}_f({\bf x})-\frac{e}{2\pi}\int
d^2x'\,{\bf B},
\end{equation}
where the integral is over a plane which is orthogonal to ${\bf B}$,
and
\begin{equation}
\widetilde{\bf M}_{EM}=e\,{\bf x}\times {\bf A}_f({\bf x}).
\end{equation}

Summing Eqs.(3), (5) and (12), one gets
\begin{equation}
{\widetilde{\bf J}}=-i\,{\bf x}\times[\bpart-ie{\bf A}({\bf
x})+ie{\bf A}_f({\bf x})]+\frac{1}{2}\bsigma.
\end{equation}
Taking into account relation
\begin{equation}
{\bf x}\times{\bf A}_f({\bf x})=\frac{1}{2\pi} \int\limits_{D}
d^2x'\,{\bf B},
\end{equation}
where the integration is over region $D$ defined by
$({x'}^1)^2+(x'^2)^2<(x^1)^2+(x^2)^2$, one can verify that
${\bf{\widetilde{J}}}$ (13) commutes with the one-particle
hamiltonian operator; thus ${\bf{\widetilde{J}}}$ (13) can be
regarded as the operator of total angular momentum of the system,
which is conserved as a consequence of rotational invariance.

If the improved quantity, Eq.(11), is taken instead of the canonical
one, Eq.(12), then, summing Eqs.(3), (5) and (11), one gets
\begin{equation}
{\bf J}=-i\,{\bf x}\times[{\bpart}-ie{\bf A}({\bf x})+ie{\bf
A}_f({\bf x})]-\frac{e}{2\pi}\int d^2x'\,{\bf B}+\frac{1}{2}\bsigma.
\end{equation}
Since the difference between operators ${\bf{\widetilde{J}}}$\, (13)
and $\bf J$\, (15) is $\bf x$-inde\-pen\-dent and diagonal in spin
indices, the latter operator commutes with the hamiltonian as well.

Although both the canonical and improved definitions of angular
momentum of electromagnetic field are compatible with the
conservation of total angular momentum, the motivation in favour of
the first one may be seen in the fact that we start initially from
the canonical angular momentum. Then, at intermediate stage, the
improved angular momentum is introduced as a convenient and
well-defined quantity which is constructed directly from the gauge
invariant and symmetric energy-momentum tensor, see Appendix A. This
quantity plays a somewhat auxiliary role allowing one to express the
canonical angular momentum through the explicitly gauge invariant
entities. Note that angular momentum ${\bf{\widetilde{J}}}$ (13) in
the gauge ${\bf A}={\bf A}_f$ coincides with angular momentum of a
free particle, Eq.(3). Thus, it is evident that spectra of operators
${\bf J}_0$ and ${\bf{\widetilde{J}}}$ coincide, and this may serve
as an additional argument in favour of the canonical definition of
angular momentum.

Note also that the difference between ${\bf J}$ (15) and
${\bf{\widetilde{J}}}$ (13) is equal to the return magnetic flux
(times $e(2\pi)^{-1}$), see Eq.(8), which completes a flux loop at
extremely large distances outside the position of the charged
particle. If the particle is allowed to go beyond the region of the
return flux, then, as it enters this region from outside and crosses
it, one arrives at expression (13) for total angular momentum inside
the region of the return flux.

In the case of a relativistic massive spinor particle, considered in
the present paper, the spin matrix is
${\bsigma}=(2i)^{-1}\balpha\times \balpha$ and the hamiltonian is
\begin{equation}
H =-i\,\balpha\cdot[\bpart-ie{\bf A}({\bf x})]+\gamma^0m,
\end{equation}
where $\balpha=\gamma^0\bgamma$, and $\gamma^0$, $\bgamma$ are the
Dirac matrices.

\section{Bohm--Aharonov configuration of the \newline magnetic field}

Of special physical interest, as it was noted in Introduction, is
the case when the region of the magnetic flux lines is not
accessible to a charged particle \cite{Aha}. Namely, one considers a
rotationally invariant magnetic field configuration in the form of a
long solenoid with extremely distant return flux, and the charged
particle is posited in the field-free region between the solenoid
and the return flux. From now on we shall call the region of
solenoid as the inner region and the field-free region as the outer
region, implying that the return flux is all pushed out to infinity.

In the outer region, improved quantity (4) is obviously vanishing,
${\bf M}_{EM}=0$, whereas canonical quantity ${\bf
\widetilde{M}}_{EM}$ is given by the same expression as Eq.(12) but
is $\bf x$-independent, since in this case it depends on the whole
flux through the inner region, ${\bf x}\times{\bf A}_f({\bf
x})=\frac{1}{2\pi}\int d^2x'{\bf B}$. Eq.(13) takes form
\begin{equation}
{\bf\widetilde{J}}=-i\,{\bf x}\times[\bpart-ie{\bf A}({\bf x})]+
\frac{e}{2\pi}\int d^2x'{\bf B}+\frac{1}{4i}\balpha\times\balpha,
\end{equation}
whereas Eq.(15) takes form of the kinetic (or mechanical) angular
momentum of the particle,
\begin{equation}
{\bf J}=-i\,{\bf x}\times[\bpart-ie{\bf A}({\bf
x})]+\frac{1}{4i}\balpha\times\balpha.
\end{equation}
Note that the outer region is not simply connected, and one may feel
free not to require the coincidence of the spectrum of total angular
momentum with that of ${\bf J}_0$ (3). Also, vanishing of the
improved angular momentum of electromagnetic field in this case may
be regarded as an argument in favour of choosing ${\bf J}$ (18) in
the capacity of total angular momentum of the system. Whether ${\bf
\widetilde{J}}$ (17) or ${\bf J}$ (18) gives the physically
meaningful angular momentum, was disputed in the literature, see,
e.g., Refs.\cite{Gold, Lip, Jac3, Wil2}. In the present paper we
shall show that namely the kinetic definition of angular momentum
provides the physically reasonable behaviour of the thermal
quadratic fluctuation and correlation in the second-quantized theory
at nonzero temperature.

But before going to the second-quantized theory, let us discuss some
manifestations of non-simply-connectedness of the outer region. Let
magnetic field $\bf B$ be directed along the $x^3$-axis in the inner
region, and the outer region correspond to $(x^1)^2+(x^2)^2>r^2_c$
with $r_c$ being the transverse size of the inner region. Then
vector potential ${\bf A}_f$ takes form
\begin{equation}
A^1_f=-\Phi\frac{x^2}{(x^1)^2+(x^2)^2},\,\,\,A^2_f=
\Phi\frac{x^1}{(x^1)^2+(x^2)^2},\,\,\,A^3_f=0,
\end{equation}
where
\begin{equation}
\Phi=\frac{1}{2\pi}\int d^2x'B^3
\end{equation}
is the total magnetic flux in units of $2\pi$, and Eq.(18) in gauge
${\bf A}={\bf A}_f$ takes form
\begin{equation}
J^3=-i(x^1\partial_2-x^2\partial_1)-e\Phi+\frac{1}{2i}\alpha^1\alpha^2.
\end{equation}
Operator (21) acts on functions which are single-valued in the outer
region:
\begin{equation}
\langle r,\,\varphi,\,x^3|=\langle r,\,\varphi+2\pi,\,x^3|,
\end{equation}
where $r$ and $\varphi$ are the polar coordinates in the
$x^1x^2$-plane. Since the outer region is not simply connected,
one can consider functions satisfying much more general condition
\begin{equation}
\langle r,\,\varphi,\,x^3|'=e^{i2\pi\Xi}\langle
r,\,\varphi+2\pi,\,x^3|',
\end{equation}
where $\Xi$ is a continuous real parameter. Functions satisfying
conditions (22) and (23) are related by gauge transformation
\begin{eqnarray}
\langle
r,\,\varphi,\,x^3|'&\!\!\!=&\!\!\!e^{-i\,\Xi\,\varphi}\langle
r,\,\varphi,\,x^3|,\nonumber \\
{\bf A}'&\!\!\!=&\!\!\!{\bf A}_f-e^{-1}\,\Xi\,\bpart \varphi.
\end{eqnarray}
Neglecting the transverse size of the inner region
$(r_c\rightarrow 0)$, and presenting the angular variable with
range $0<\varphi<2\pi$ as
\begin{equation}
\varphi=\arctan\left(\frac{x^2}{x^1}\right)+\pi\left[\theta(-x^1)
+2\theta(x^1)\,\,\theta(-x^2)\right],
\end{equation}
where $\theta(u)=\left\{\begin{array}{cc}
  1, & u>0 \\
  0, & u<0
\end{array}\right\}$, one gets immediately
\begin{equation}
\partial_1\varphi=-\frac{x^2}{(x^1)^2+(x^2)^2},\,\,\,\partial_2\varphi
=\frac{x^1}{(x^1)^2+(x^2)^2}-2\pi\theta(x^1)\delta(x^2),
\end{equation}
and, consequently,
\begin{eqnarray}
\!\!\!{A'}^1&\!\!\!=&\!\!\!-(\Phi-e^{-1}\Xi)\frac{x^2}{(x^1)^2+(x^2)^2},\nonumber \\
\!\!\!{A'}^2&\!\!\!=&\!\!\!(\Phi-e^{-1}\Xi)\frac{x^1}{(x^1)^2+(x^2)^2}+2\pi
e^{-1}\Xi\,\theta(x^1)\delta(x^2), \\
\!\!\!{A'}^3&\!\!\!=&\!\!\!0. \nonumber
\end{eqnarray}
Although the curl of the vector potential remains invariant under
gauge transformation (24),
\begin{equation}
B^3=\partial_1A^2_f-\partial_2A^1_f=\partial_1{A'}^2-\partial_2{A'}^1=
2\pi\Phi\,\delta(x^1)\delta(x^2),
\end{equation}
angular momentum (21) is changed to
\begin{equation}
J'^3=-i(x^1\partial_2-x^2\partial_1)-e\Phi+\Xi+\frac{1}{2i}\alpha^1\alpha^2,
\end{equation}
but its spectrum remains unchanged, because operator (29) acts on
functions satisfying condition (23). In particular, choosing
$\Xi=e\Phi$ one gets the gauge with the vector potential eliminated
everywhere excepting the semiaxis of positive $x^1$:
\begin{equation}
{A'}^1=0,\,\,\,{A'}^2=2\pi\Phi\,
\theta(x^1)\delta(x^2),\,\,\,\,{A'}^3=0.
\end{equation}
Kinetic angular momentum (18) in this gauge coincides with angular
momentum of a free particle, Eq.(3), but their spectra differ: the
operator of the latter acts on functions which are defined
everywhere and, thus, single-valued (22), while the operator of the
former acts on functions which are defined in space with a cut along
the positive $x^1$ semiaxis and, thus, satisfying boundary condition
(23) with $\Xi=e\Phi$ on the sides of the cut. The cut can be
continuously deformed, and, in general, its projection on the
$x^1x^2$-plane can be a curved line starting from the origin and
going to infinity. Eq.(30) in the case of such cut takes form
$$
{A'}^j=-2\pi\Phi\int\limits_{0}^{\infty}\varepsilon^{3jj'}\delta^2({\bf
x}- {\bf x}')dx'^{j'},
$$
where coordinate vectors $\bf x$ and ${\bf x}'$ lie in the
$x^1x^2$-plane.

To conclude this section, we note that, obviously, canonical angular
momentum (17) in gauge ${\bf A}={\bf A}_f$ takes form
\begin{equation}
\tilde{J}^3=-i(x^1\partial_2-x^2\partial_1)+\frac{1}{2i}\alpha^1\alpha^2,
\end{equation}
and it is changed to
\begin{equation}
\tilde{J}'^3=-i(x^1\partial_2-x^2\partial_1)+\Xi+\frac{1}{2i}\alpha^1\alpha^2
\end{equation}
under gauge transformation (24). The spectrum of canonical angular
momentum is given by Eq.(1), while the spectrum of kinetic angular
momentum is given by Eq.(2).

\section{Second-quantized theory at finite \newline temperature}

The operator of the second-quantized fermion field in a static
background can be presented in the form
\begin{equation}\label{intr1}
\Psi(\textbf{x},
t)=\sum\hspace{-1.7em}\int\limits_{(E_\lambda>0)}e^{-iE_\lambda
t}\langle\textbf{x}| \lambda\rangle a_{\lambda}+
\sum\hspace{-1.7em}\int\limits_{(E_\lambda<0)}e^{-iE_\lambda
t}\langle\textbf{x}| \lambda\rangle b^+_{\lambda}\,,
\end{equation}
where $a^+_{\lambda}$ and $a_{\lambda}$ $(b^+_{\lambda}$ and
$b_{\lambda})$ are the fermion (antifermion) creation and
destruction  operators satisfying anticommutation relations,
\begin{equation}\label{intr2}
\left[a_{\lambda},a^+_{\lambda'}\right]_+=
\left[b_{\lambda},b^+_{\lambda'}\right]_+= \langle
\lambda|\lambda'\rangle\,,
\end{equation}
and $\langle \textbf{x}|\lambda\rangle$ is the solution to the
stationary Dirac equation,
\begin{equation}\label{intr3}
H\langle \textbf{x}|\lambda\rangle=E_\lambda\langle
\textbf{x}|\lambda\rangle\,,
\end{equation}
$H$ is the Dirac (one-particle) hamiltonian, $\lambda$ is the set of
parameters (quantum numbers) specifying a one-particle state, and
$E_\lambda$ is the energy of the state; symbol
${\displaystyle\sum\hspace{-1.4em}\int\,}$ means the summation over
discrete and the integration (with a certain measure) over
continuous values of $\lambda$.  Ground state $|{\rm vac}\rangle$ of
the second-quantized theory is defined as
\begin{equation}\label{intr4}
a_\lambda|{\rm vac}\rangle=b_\lambda|{\rm vac}\rangle=0\,.
\end{equation}

Let $J$ be an operator commuting with the hamiltonian in the
first-quantized theory,
\begin{equation}\label{intr5}
  [J,H]_-=0\,.
\end{equation}
In the case of unbounded operators, commutation of their
resolvents is implied, or, to be more specific, it is sufficient
to require that operators $H$ and $J$ have a common set of
eigenfunctions, i.e. relation
\begin{equation}\label{intr6}
J\langle\textbf{x}|\lambda\rangle=j_\lambda\langle\textbf{x}|\lambda\rangle
\end{equation}
holds as well as Eq.(35). Eigenfunctions
$\langle\textbf{x}|\lambda\rangle$ satisfy the conditions of
completeness and orthonormality; in general, normalization to a
delta function is implied. Thus, in the second-quantized theory, the
operators of the dynamical variables (physical observables)
corresponding to $H$ and $J$ can be diagonalized:
\begin{equation}\label{intr7}
{\hat U}\equiv\frac12\int d^d
x\left[\Psi^+(\textbf{x},t),\,H\Psi(\textbf{x},t)\right]_-
=\sum\hspace{-1.3em}\int \,\,E_\lambda\left[a^+_\lambda
a_\lambda-b^+_\lambda b_\lambda-\frac12{\rm sgn}(E_\lambda)\right]
\end{equation}
and
\begin{equation}\label{intr8}
\hat M\equiv\frac12 \int d^d
x\left[\Psi^+(\textbf{x},t),\,J\Psi(\textbf{x},t)\right]_-
=\sum\hspace{-1.3em}\int\,\, j_\lambda\left[a^+_\lambda
a_\lambda-b^+_\lambda b_\lambda-\frac12{\rm
sgn}(E_\lambda)\right]\,,
\end{equation}
$d$ is the space dimension, and ${\rm sgn}(u)=\theta(u)-\theta(-u)$
is the sign function.

The thermal average of the observable corresponding to operator (40)
is conventionally defined as (see, e.g., Ref.\cite{Das})
\begin{equation}
M(T)=\langle\hat{M}\rangle_\beta\equiv\frac{Sp\, \hat{M}e^{-\beta
\hat{U}}}{ Sp\,e^{-\beta \hat{U}}},\,\,\,\, \beta=(k_B T)^{-1},
\end{equation}
where $T$ is the equilibrium temperature, $k_B$ is the Boltzmann
constant, and $Sp$ is the trace or the sum over the expectation
values in the Fock state basis created by operators in Eq.(34). Also
one defines the thermal quadratic fluctuation
\begin{equation}
\Delta(T;\hat{M},\hat{M})=\langle\hat{M}^2\rangle_\beta-
(\langle\hat{M}\rangle_\beta)^2.
\end{equation}
Quantities (41) and (42) can be expressed through the derivatives of
the appropriate thermodynamic potential:
\begin{equation}
M(T)=-\frac{\partial\Omega(\beta,\mu)}{\partial\mu}\biggr|_{\mu=0},
\,\,\Delta(T;\hat{M},\hat{M})=-\frac{1}{\beta}
\frac{\partial^2\Omega(\beta,\mu)}{\partial\mu^2}\biggr|_{\mu=0},
\end{equation}
where $\mu$ is the appropriate chemical potential, and
\begin{equation}
\Omega(\beta,\mu)=-\frac{1}{\beta}\ln
Sp\exp[-\beta(\hat{U}-\mu\hat{M})].
\end{equation}
We show in Appendix B that thermodynamic potential (44) is presented
as
\begin{equation}
\Omega(\beta,\mu)=-\frac{1}{\beta}{\rm Tr}\ln
\cosh[\frac{1}{2}\beta(H-\mu J)],
\end{equation}
where ${\rm Tr}$ is the trace of an integro-differential operator in
the functional space: ${\rm Tr}\ldots =\int d^d x\,{\rm
tr}\langle{\bf x}|\ldots |{\bf x}\rangle$; ${\rm tr}$ denotes the
trace over spinor indices only. Then, using Eq.(43), one can express
average (41) and fluctuation (42) through functional traces of
operators in the first-quantized theory:
\begin{equation}
M(T)=-\frac{1}{2}{\rm Tr}J\tanh (\frac{1}{2}\beta H)
\end{equation}
and
\begin{equation}
\Delta(T;\hat{M},\hat{M})=\frac{1}{4}{\rm Tr}J^2\, {\rm
sech}^2(\frac{1}{2}\beta H).
\end{equation}
Eqs.(46) and (47) are transformed into integrals over the energy
spectrum,
\begin{equation}
M(T)=-\frac{1}{2}\int\limits_{-\infty}^{\infty}dE\,\tau_J(E)
\tanh(\frac{1}{2}\beta E)
\end{equation}
and
\begin{equation}
\Delta(T;\hat{M},\hat{M})=\frac{1}{4}
\int\limits_{-\infty}^{\infty}dE\,\tau_{J^2}(E)\,{\rm
sech}^2(\frac{1}{2}\beta E),
\end{equation}
where
\begin{equation}
\tau_J(E)={\rm Tr}\,J\,\delta(H-E)
\end{equation}
and
\begin{equation}
\tau_{J^2}(E)={\rm Tr}\,J^2\,\delta(H-E)
\end{equation}
are the appropriate spectral densities.

Note that in the case of $J=I$, where $I$ is the unit matrix in the
space of Dirac matrices, the corresponding operator in the
second-quantized theory is the operator of fermion number; $\mu$ and
$\Omega$ are the conventional chemical and thermodynamic potentials
in this case. In the $d=1$ case fermion number is the only
observable which is conserved in addition to energy. In more than
one dimensions there are more conserved observables. In particular,
in the $d=2$ case, in addition to energy and fermion number, also
total angular momentum is conserved when the system is rotationally
invariant, as it was discussed in two previous sections.

If an observable is not conserved, then its operator in the
first-quantized theory does not commute with hamiltonian,
$[\Upsilon,\,H]\neq 0$, and its operator in the second-quantized
theory,
\begin{equation}
\hat{O}=\frac{1}{2}\int d^d x\left[\Psi^+({\bf
x},\,t),\,\Upsilon\Psi({\bf x},\,t)\right]_-,
\end{equation}
is not diagonalizable. Similarly to Eq.(48), the thermal average of
the nonconserved observable can be presented as
\begin{equation}
O(T)=-\frac{1}{2}\int\limits_{-\infty}^{\infty}
dE\,\tau_\Upsilon(E)\tanh(\frac{1}{2}\beta E),
\end{equation}
where
\begin{equation}
\tau_\Upsilon(E)={\rm Tr}\Upsilon\,\delta(H-E).
\end{equation}
Also one can consider the thermal correlation of the conserved and
nonconserved observables,
\begin{equation}
\Delta(T;\,\hat{O},\,\hat{M})=\langle\hat{O}\,\hat{M}\rangle_\beta
-\langle\hat{O}\rangle_\beta\langle\hat{M}\rangle_\beta,
\end{equation}
which, similarly to Eq.(49), can be presented as
\begin{equation}
\Delta(T;\,\hat{O},\,\hat{M})=\frac{1}{4}\int\limits_{-\infty}^{\infty}dE\,
\tau_{\Upsilon J}(E){\rm sech}^2(\frac{1}{2}\beta E),
\end{equation}
where
\begin{equation}
\tau_{\Upsilon J}(E)={\rm Tr}\Upsilon J\,\delta(H-E).
\end{equation}

It should be noted that relations (45), (48), (49), (53), and (56)
are somewhat formal, and the proper treatment of the normal ordering
of operator product in the second-quantized theory is required. In
the absence of interaction, the operators are normal ordered (see,
e.g., Ref.\cite{Itzyk}), i.e. the $c$-number pieces in Eqs.(39) and
(40) are dropped. Consequently, in the presence of interaction with
external fields, the $c$-number pieces in Eqs.(39) and (40) are
renormalized by subtracting these dropped pieces. Correspondingly,
thermodynamic potential (45) is decomposed as
\begin{equation}
\Omega(\beta,\mu)=\Omega^{(0)}(\beta,\mu)+\Omega^{(1)}(\beta,\mu),
\end{equation}
where
\begin{equation}
\Omega^{(0)}(\beta,\mu)=-\frac{1}{\beta}{\rm Tr}\ln\left\{1+\exp
\left[-\beta\left(|H_0|-\mu J_0\,{\rm
sgn}(H_0)\right)\right]\right\}
\end{equation}
is the thermodynamic potential in the free field case, and
\begin{equation}
\Omega^{(1)}(\beta,\mu)=-\frac{1}{\beta} \left\{{\rm Tr}\ln {\rm
cosh}\left[\frac{1}{2}\beta(H\!-\!\mu J)\right]\!-\!{\rm
Tr}\ln\cosh \left[\frac{1}{2}\beta(H_0\!-\!\mu J_0)\right]\right\}
\end{equation}
is the addition which is due to interaction with external fields;
here the operators in the first-quantized theory without interaction
are denoted by $H_0$ and $J_0$. Note that term $-\frac{1}{2}{\rm
Tr}\left[|H_0|-\mu J_0{\rm sgn}(H_0)\right]$ is dropped, which
corresponds to the normal ordering at zero temperature in the free
field case.

As a consequence of Eqs.(58)-(60) one gets
\begin{equation}
M(T)=M^{(0)}(T)+M^{(1)}(T)
\end{equation}
and
\begin{equation}
\Delta(T;\hat{M},\hat{M})=\Delta^{(0)}(T;\hat{M},
\hat{M})+\Delta^{(1)}(T;\hat{M},\hat{M}),
\end{equation}
where
\begin{equation}
M^{(0)}(T)=\int\limits_{-\infty}^{\infty}dE\,\tau^{(0)}_J(E)
\,\frac{{\rm sgn} (E)}{e^{\beta |E|}+1},
\end{equation}
\begin{equation}
M^{(1)}(T)=-\frac{1}{2}\int\limits_{-\infty}^{\infty}
dE\,\tau^{(1)}_J(E)\tanh(\frac{1}{2}\beta E),
\end{equation}
\begin{equation}
\tau^{(0)}_J(E)={\rm Tr}J_0\,\delta(H_0-E),\,\,\tau^{(1)}_J(E)= {\rm
Tr}J\,\delta(H-E)-{\rm Tr}J_0\,\delta(H_0-E),
\end{equation}
and
\begin{equation}
\Delta^{(0)}(T;\hat{M},\hat{M})=\frac{1}{4}
\int\limits_{-\infty}^{\infty}dE\,\tau^{(0)}_{J^2}(E)\, {\rm sech}^2
(\frac{1}{2}\beta E),
\end{equation}
\begin{equation}
\Delta^{(1)}(T;\hat{M},\hat{M})=\frac{1}{4}\int
\limits_{-\infty}^{\infty}dE\,\tau^{(1)}_{J^2}(E)\,{\rm
sech}^2(\frac{1}{2}\beta E),
\end{equation}
\begin{equation}
\tau^{(0)}_{J^2}(E)={\rm
Tr}J_0^2\,\delta(H_0-E),\,\,\tau^{(1)}_{J^2}(E)={\rm
Tr}J^2\,\delta(H\!-\!E)-{\rm Tr}J^2_0\,\delta(H_0-E).
\end{equation}

Using relation $\delta(H-E)=\frac{1}{2\pi
i}\left[(H-E-i0)^{-1}-(H-E+i0)^{-1}\right]$, one can transform
integrals over the real energy spectrum in Eqs.(64) and (67) into
integrals over a contour on the complex energy plane, thus yielding
a representation of thermal characteristics through the renormalized
resolvent traces,
\begin{equation}
M^{(1)}(T)=-\frac{1}{2}\int\limits_{C}\frac{d\omega}{2\pi i}
\left[{\rm Tr}J(H-\omega)^{-1}\right]_{\rm
ren}\tanh(\frac{1}{2}\beta\omega)
\end{equation}
and
\begin{equation}
\Delta^{(1)}(T;\hat{M},\hat{M})=\frac{1}{4}\int\limits_{C}
\frac{d\omega}{2\pi i}\left[{\rm Tr}J^2(H-\omega)^{-1}\right]_{\rm
ren}{\rm sech}^2(\frac{1}{2}\beta\omega),
\end{equation}
where $C$ is the contour consisting of two collinear straight lines,
$(-\infty+i0,\,+\infty+i0)$ and $(+\infty-i0,\,-\infty-i0)$, in the
complex $\omega$-plane, and the renormalized resolvent traces are
\begin{equation}
\left[{\rm Tr}J(H-\omega)^{-1}\right]_{\rm ren}={\rm
Tr}J(H-\omega)^{-1}-{\rm Tr}J_0(H_0-\omega)^{-1}
\end{equation}
and
\begin{equation}
\left[{\rm Tr}J^2(H-\omega)^{-1}\right]_{\rm ren}={\rm
Tr}J^2(H-\omega)^{-1}-{\rm Tr}J_0^2(H_0-\omega)^{-1}.
\end{equation}

To conclude this section, we note that thermal average (53) and
correlation (56) can be treated in a similar way.

\section{Traces of resolvents and spectral densities}

Let us consider quantization of the spinor field on a plane ($d=2$)
which is orthogonal to the Bohm--Aharonov magnetic field
configuration. Dirac hamiltonian (16) in gauge ${\bf A}={\bf A}_f$,
where ${\bf A}_f$ is given by Eq.(19), takes form
\begin{equation}
H=-i\alpha^r\partial_r-ir^{-1}\alpha^\varphi(\partial\varphi-ie\Phi)+\gamma^0
m,
\end{equation}
where $\Phi$ is the total magnetic flux of the solenoid
(Bohm--Aharonov vortex), see Eq.(20), and
$$
\alpha^r=\alpha^1\cos \varphi+\alpha^2\sin \varphi,\,\,\,\,
\alpha^\varphi=-\alpha^1\sin\varphi+\alpha^2\cos \varphi.
$$
In $2+1$-dimensional space-time the Clifford algebra has two
inequivalent irreducible representations which can be differed in
the following way:
\begin{equation}
\alpha^1\alpha^2\gamma^0=is,\,\,\,\,s=\pm1.
\end{equation}
Choosing the $\gamma_0$ matrix in the diagonal form
\begin{equation}
\gamma^0=\sigma_3,
\end{equation}
one gets
\begin{equation}
\alpha^1=-e^{\frac{i}{2}\sigma_3\chi_s}\sigma_2e^{-\frac{i}{2}\sigma_3\chi_s},\,\,
\alpha^2=s
e^{\frac{i}{2}\sigma_3\chi_s}\sigma_1e^{-\frac{i}{2}\sigma_3\chi_s},
\end{equation}
where $\sigma_1$, $\sigma_2$, and $\sigma_3$ are the Pauli matrices,
and $\chi_1$ and $\chi_{-1}$ are the parameters varying in interval
$0<\chi_s<2\pi$ to go over to equivalent representations. Note also
that in odd-dimensional space-time the $m$ parameter in Eq.(16) can
take both positive and negative values; a change of sign of $m$
corresponds to going over to the inequivalent representation.

In view of Eq.(74), the kinetic, Eq.(21), and canonical, Eq.(31),
angular momenta of the planar system take form
\begin{equation}
J=-i\partial_\varphi-e\Phi+\frac{1}{2}s\gamma^0
\end{equation}
and
\begin{equation}
\widetilde{J}=-i\partial_\varphi+\frac{1}{2}s\gamma^0,
\end{equation}
respectively.

The kernel of the resolvent (the Green's function) of the Dirac
hamiltonian in the coordinate representation is defined as
\begin{equation}\label{c1}
G^\omega(r,\varphi;r',\varphi')=\langle
r,\varphi|(H-\omega)^{-1}|r',\varphi' \rangle,
\end{equation}
where $\omega$ is a complex parameter with dimension of energy.
Taking into account Eqs.(75) and (76), one expands Eq.(79) in modes
and gets the following expression
\begin{equation}\label{c2}
G^\omega(r,\varphi;r',\varphi')\!=\!\frac{1}{2\pi}\sum\limits_{n=-\infty}^{\infty}
e^{in(\varphi-\varphi')} \!\left(\!\!\begin{array}{cc}
a_n(r;r')&\!\!
d_n(r;r')e^{-i(s\varphi'-\chi_s)}\\
b_n(r;r')e^{i(s\varphi-\chi_s)} &\!\!
c_n(r;r')e^{is(\varphi-\varphi')}\end{array}\!\!\right).
\end{equation}
In the case of $H$ given by Eq.(73) radial components of
$G^\omega(r,\varphi;r',\varphi')$ (\ref{c2}) satisfy equations
%\begin{multline}
$$
\left(\!\!\begin{array}{cc}
-\omega+m &\partial_r+s(n-e\Phi+s)r^{-1}\\
-\partial_r+s(n-e\Phi)r^{-1}&-\omega-m\end{array}\!\!\right)
\left(\!\!\begin{array}{cc}
a_n(r;r')& d_n(r;r')  \\
b_n(r;r') & c_n(r;r')\end{array}\!\!\right)\!\!=
$$
$$
=\left(\!\!\begin{array}{cc}
-\omega+m &\partial_{r'}+s(n-e\Phi+s){r'}^{-1}\\
-\partial_{r'}+s(n-e\Phi){r'}^{-1}&-\omega-m\end{array}\!\!\right)
\left(\begin{array}{cc} a_n(r;r')& b_n(r;r')\\
d_n(r;r')& c_n(r;r')\end{array}\right)=
$$
\begin{equation}
=\frac{\delta(r-r')}{\sqrt{rr'}}
\left(\begin{array}{cc} 1& 0\\ 0 & 1\end{array}\right).
\end{equation}
%\end{multline}
All radial components behave asymptotically at large distances
($r\rightarrow\infty$ or $r'\rightarrow\infty$) as outgoing waves
($e^{ikr}/(2\pi\sqrt{r})$ or $e^{ikr'}/(2\pi\sqrt{r'})$); here
$k=\sqrt{\omega^2-m^2}$ and a physical sheet for square root is
chosen as $0<{\rm Arg}\,k<\pi$ (${\rm Im}\,k>0$). Another boundary
condition should be imposed at the boundary of the inner region
containing a magnetic vortex. In the absence of the vortex,
$e\Phi=0$, all radial components obey the condition of regularity at
small distances ($r\rightarrow 0$ or $r'\rightarrow 0$), and this
corresponds to the fact that free hamiltonian $H_0$ (i.e. $H$ (73)
at $e\Phi=0$) is essentially self-adjoint. In the presence of the
vortex, $e\Phi\neq 0$, when a size of the inner region is neglected
($r_c\rightarrow 0$), the condition of regularity at small distances
cannot be imposed on all radial components. This is due to the fact
that hamiltonian $H$ (73) is not essentially self-adjoint, and a
parameter of the boundary condition at the location of the vortex
(at the origin) exhibits itself as a parameter of the self-adjoint
extension of the hamiltonian operator, for details see
Refs.\cite{Ger, Si6, Si7}. To be more precise, partial hamiltonians
are essentially self-adjoint for all $n$ with the exception of
$n=n_c$, where
\begin{equation}
n_c=[\![e\Phi]\!]+\frac{1}{2}-\frac{1}{2}s,
\end{equation}
$[\![u]\!]$ is the integer part of quantity $u$ (i.e., the largest
integer which is less than or equal to $u$).  The partial
Hamiltonian for $n=n_c$ requires a self-adjoint extension according
to the Weyl -- von Neumann theory of self-adjoint operators (see,
e.g., Ref.\cite{Alb}). Appropriately, radial components $a_n$,
$b_n$, $c_n$, and $d_n$ in Eq.(80) with $n\neq n_c$ are regular at
$r\rightarrow 0$ and $r'\rightarrow 0$, whereas those with $n=n_c$
satisfy conditions (for details see Ref.\cite{Si4}):
\begin{equation}\label{c8}
\left.\begin{array}{l}
\!\!{\displaystyle\cos\left(s\frac\Theta2+\frac\pi4\right)\lim_{r\rightarrow0}
(|m|r)^Fa_{n_c}(r;r')}=
{\displaystyle-{\rm{sgn}}(m)\sin\left(s\frac\Theta2+\frac\pi4\right)
\lim_{r\rightarrow0}(|m|r)^{1-F}b_{n_c}(r;r')}\vspace{0.3em}\\
\!\!{\displaystyle\cos\left(s\frac\Theta2+\frac\pi4\right)
\lim_{r\rightarrow0}(|m|r)^Fd_{n_c}(r;r')}=
{\displaystyle-{\rm{sgn}}(m)\sin\left(s\frac\Theta2+\frac\pi4\right)
\lim_{r\rightarrow0}(|m|r)^{1-F}c_{n_c}(r;r')}
\end{array}\!\!\right\},
\end{equation}
and
\begin{equation}\label{c9}
\left.\begin{array}{l}
\!\!\!{\displaystyle\cos\!\left(\!s\frac\Theta2+\frac\pi4\right)
\lim_{r'\rightarrow0}(|m|r')^Fa_{n_c}(r;r')}=
{\displaystyle-{\rm{sgn}}(m)\sin\!\left(\!s\frac\Theta2+
\frac\pi4\right)\lim_{r'\rightarrow0}(|m|r')^{1-F}d_{n_c}(r;r')}\vspace{0.3em}\\
\!\!\!{\displaystyle\cos\!\left(\!s\frac\Theta2+\frac\pi4\right)
\lim_{r'\rightarrow0}(|m|r')^Fb_{n_c}(r;r')}=
{\displaystyle-{\rm{sgn}}(m)\sin\!\left(\!s\frac\Theta2+
\frac\pi4\right)\lim_{r'\rightarrow0}(|m|r')^{1-F}c_{n_c}(r;r')}
\end{array}\!\!\right\},
\end{equation}
where $\Theta$ is the self-adjoint extension parameter, and
\begin{equation}\label{c10}
F=s(e\Phi - [\![e\Phi]\!])+\frac12-\frac12s\,;
\end{equation}
note here that Eqs.(83) and (84) imply that $0<F<1$, since in the
case of ${\displaystyle F=\frac12-\frac12s}$ all radial components
obey the condition of regularity at $r\rightarrow0$ and
$r'\rightarrow0$. Note also that Eqs.(83) and (84) are periodic in
$\Theta$ with period $2\pi$.

The radial components of the resolvent kernel in the presence and in
the absence of the vortex are listed in Appendix C.

Let us consider quantities

\begin{equation}\label{c12}
tr \, G^\omega(r,\varphi;r',\varphi)=
\!\frac{1}{2\pi}\sum_{n=-\infty}^\infty[a_n(r;r')+c_n(r;r')]\,,
\end{equation}

\begin{equation}\label{c14}
tr \, J\, G^\omega(r,\varphi;r',\varphi)=
\!\frac{1}{2\pi}\sum_{n=-\infty}^\infty\left(n-e\Phi+\frac12
s\right)[a_n(r;r')+c_n(r;r')]\,,
\end{equation}

\begin{equation}\label{c15}
tr \, J^2\, G^\omega(r,\varphi;r',\varphi)=
\!\frac{1}{2\pi}\sum_{n=-\infty}^\infty\left(n-e\Phi+\frac12
s\right)^2[a_n(r;r')+c_n(r;r')]\,.
\end{equation}
In the absence of the vortex, using Eqs.(C.14)-(C.17) and performing
summation over $n$, one gets in the case of $Im\,k>|{\rm Re}\,k|$:
\begin{equation}
tr\, G_0^\omega(r,\varphi;\,r',\varphi)=
\frac{\omega}{\pi}K_0(-ik|r-r'|),
\end{equation}
\begin{equation}
tr \, J_0\, G_0^\omega(r,\varphi;\,r',\varphi)=
\frac{sm}{2\pi}K_0(-ik|r-r'|),
\end{equation}
\begin{equation}
tr \, J_0^2 \,G_0^\omega(r,\varphi;\,r',\varphi)=
\frac{\omega}{4\pi}\left[K_0(-ik|r-r'|)-
\frac{4ikrr'}{|r-r'|}K_1(-ik|r-r'|)\right],
\end{equation}
where $K_\rho(u)$ is the Macdonald function of order $\rho$. In the
presence of the vortex, quantities (86)-(88) consist of two pieces:
one is finite in the limit $r'\rightarrow r$, and another one, which
is divergent in this limit, coincides with quantities (89)-(91),
correspondingly. Moreover, the difference between appropriate
quantities in the presence and in the absence of the vortex at
$r'=r$ is exponentially decreasing as $r\rightarrow \infty$, and
this allows us to perform integration over the twodimensional
infinite spatial volume,
$\int\limits_{0}^{2\pi}d\varphi\int\limits_{0}^{\infty}dr\,r$, and
obtain the renormalized traces (see Refs.\cite{Si4, Si5}):
\begin{equation}
\left[Tr\,(H\!-\!\omega)^{-1}\right]_{\rm
ren}\!=-\frac{1}{\omega^2-m^2} \left[\frac{F(\omega+m)\tan
\nu_\omega+(1-F)(\omega-m)e^{iF\pi}}{\tan
\nu_\omega+e^{iF\pi}}-F(1\!-\!F)\omega\right],
\end{equation}
\begin{equation}
\left[Tr\,J\,(H\!-\!\omega)^{-1}\right]_{\rm
ren}\!=-s(F-\frac{1}{2})[Tr\,(H-\omega)^{-1}]_{\rm ren}+
\frac{sF(1\!-\!F)}{\omega^2-m^2}\left[\frac{1}{3}(F-\frac{1}{2})\omega
+\frac{1}{2}m\right],
\end{equation}
\begin{equation}
\left[Tr\,J^2\,(H\!-\!\omega)^{-1}\right]_{\rm
ren}\!=(F-\frac{1}{2})^2[Tr\,(H-\omega)^{-1}]_{\rm ren}+
\frac{F(1\!-\!F)}{\omega^2\!-\!m^2}\left[\frac{1}{2}F(1\!-\!F)\omega-
\frac{2}{3}(F\!-\!\frac{1}{2})m\right]\!;
\end{equation}
here $\tan \nu_\omega$ is given by Eq.(C.13), and results (92)-(94)
are analytically continued from region $Im\,k>|{\rm Re}\,k|$ to
region $Im\,k>0$, i.e. to the whole complex $\omega$-plane.

To deal with the divergent at $r'\rightarrow r$ pieces of the
resolvent kernels, it is sufficient to regularize the resolvent
kernel in the absence of the vortex and define
\begin{equation}
G_0^{\omega,t}(r,\varphi;\,r',\varphi')=\left\langle r,\varphi
\left|(H_0-\omega)^{-1}\exp(-tH_0^2)\right|r',\varphi'\right\rangle,
\end{equation}
where $t>0$ is the regularization parameter. In Appendix D we obtain
the following relations in the case of $Im\,k>|{\rm Re\,k}|$:
\begin{equation}
tr \, G_0^{\omega,t}(r,\varphi;\,r,\varphi)=\frac{\omega}{2\pi}
e^{-t(m^2+k^2)}E_1(-tk^2),
\end{equation}
\begin{equation}
tr\,J_0\,G_0^{\omega,t}(r,\varphi;\,r,\varphi)=\frac{sm}{4\pi}
e^{-t(m^2+k^2)}E_1(-tk^2),
\end{equation}
\begin{equation}
tr\,J_0^2\,G_0^{\omega,t}(r,\varphi;\,r,\varphi)=\frac{\omega}{4\pi}
e^{-tm^2}\left[r^2t^{-1}+(r^2k^2+\frac{1}{2})e^{-tk^2}E_1(-tk^2)\right],
\end{equation}
where
$$
E_1(u)=\int\limits_{u}^{\infty}\frac{du}{u}e^{-u}
$$
is the exponential integral (see, e.g., Ref.\cite{Abra}); note that
Eqs.(96)-(98) are analytically continued to region $Im\,k>0$, i.e.
to the whole complex $\omega$-plane. Integrating over the
twodimensional spatial volume,
$\int\limits_{0}^{2\pi}d\varphi\int\limits_{0}^{R}dr\,r$, where $R$
is the volume radius, we get
\begin{equation}
Tr\,(H_0-\omega)^{-1}e^{-tH_0^2}=\frac{1}{2}R^2\omega
\,e^{-t\omega^2} E_1[t(m^2-\omega^2)],
\end{equation}
\begin{equation}
Tr\,J_0\,(H_0-\omega)^{-1}e^{-tH_0^2}=\frac{1}{4}R^2sm
\,e^{-t\omega^2} E_1[t(m^2-\omega^2)],
\end{equation}
\begin{equation}
Tr\,J_0^2\,(H_0-\omega)^{-1}e^{-tH_0^2}=\frac{1}{8}R^2\omega
\left\{R^2t^{-1} e^{-tm^2}+[R^2(\omega^2-m^2)+1]e^{-t\omega^2}
E_1[t(m^2-\omega^2)]\right\}.
\end{equation}
Although traces (99)-(101) are divergent in the limit $t\rightarrow
0_{+}$, the divergences do not contribute to the physical
quantities, e.g., Eqs.(63) and (66). This is due to a specific form
of a discontinuity of the exponential integral at negative real
values of its argument, $Im\,E_1(-u\mp i0)=\pm i\pi$ ($u>0$).
Consequently, we get finite spectral densities
\begin{equation}
\tau_I^{(0)}(E)=\pm\lim\limits_{t\rightarrow 0_+}\frac{1}{\pi}Im{\rm
Tr}(H_0-E\mp i0)^{-1}e^{-tH_0^2}=\frac{1}{2}R^2|E|\,\theta(E^2-m^2),
\end{equation}
\begin{equation}
\tau_J^{(0)}(E)=\pm\lim\limits_{t\rightarrow 0_+}\frac{1}{\pi}Im{\rm
Tr}J_0(H_0-E\mp i0)^{-1}e^{-tH_0^2}=\frac{1}{4}R^2 s m\,{\rm
sgn}(E)\,\theta(E^2-m^2),
\end{equation}
\begin{equation}
\tau_{J^2}^{(0)}(E)=\pm\lim\limits_{t\rightarrow
0_+}\frac{1}{\pi}Im{\rm Tr}J_0^2(H_0-E\mp
i0)^{-1}e^{-tH_0^2}=\frac{1}{8}R^2|E|[R^2(E^2-m^2)+1]
\,\theta(E^2-m^2).
\end{equation}

\section{Thermal averages, fluctuations, and correlations}

As follows from two preceding sections, the expressions for thermal
characteristics consist of two pieces, see Eqs.(61) and (62): the
one, denoted by superscript $^{(0)}$ and corresponding to the ideal
gas contribution, depends essentially on the size of the system (R),
see Eqs.(102)-(104), and the other one, denoted by superscript
$^{(1)}$ and corresponding to the correction due to interaction with
a magnetic vortex, is independent of $R$ in the limit
$R\rightarrow\infty$, see Eqs.(92)-(94). However, it appears that
the ideal gas contribution to some characteristics is vanishing, and
these characteristics are finite as the size of the system
increases.

In particular, this is the case for the average fermion number of
the system. Let us denote the operator of fermion number by
$\hat{N}$, then it is given by Eq.(40) with $\hat{N}$, $I$ and 1
substituted for $\hat{M}$, $J$ and $j_\lambda$, respectively. Taking
into account Eq.(102), one gets $N^{(0)}(T)=0$, and using Eq.(92),
one gets (see Ref.\cite{Si4}):

%\begin{multline}\label{d9}
\begin{eqnarray}
N(T)=-\frac{\sin(F\pi)}{\pi}\int\limits_0^\infty\frac{du}
{u\sqrt{u+1}}\tanh\left(\frac12\beta m\sqrt{u+1}\right)\times
\nonumber\\ \times
\frac{Fu^FA-(1-F)u^{1-F}A^{-1}+u\left[\left(F-\frac12\right)
(u^FA+u^{1-F}A^{-1})-\cos(F\pi)\right]}
{[u^FA-u^{1-F}A^{-1}+2\cos(F\pi)]^2+4(u+1)\sin^2(F\pi)}\,-\nonumber\\
- \frac12\theta(-\cos \Theta)\tanh\left(\frac12\beta E_{BS}\right),
\end{eqnarray}
%\end{multline}
where
\begin{equation}\label{d5}
A=2^{1-2F}\frac{\Gamma(1-F)}{\Gamma(F)}\tan\left(s\frac\Theta2+\frac\pi4\right)\,,
\end{equation}
$\Gamma(u)$ is the Euler gamma function, $E_{BS}$ is the energy of
the bound state in the one-particle spectrum, which is determined as
a real root of algebraic equation (for details see Ref.\cite{Si7})
\begin{equation}\label{d6}
\frac{(1-m^{-1}E_{BS})^F}{(1+m^{-1}E_{BS})^{1-F}}\,A=-1\,;
\end{equation}
note that the bound state exists at $\cos \Theta<0$ $(A<0)$, and its
energy is zero at $A=-1$, and, otherwise, one has $0<|E_{BS}|<|m|$
and
\begin{equation}\label{d7}
{\rm{sgn}}(E_{BS})=\frac12\,{\rm{sgn}}(m)[\,{\rm{sgn}}(1+A^{-1})-{\rm{sgn}}(1+A)]\,.
\end{equation}
Contrary to the average, the quadratic fluctuation of fermion number
consists of two contributions. Using Eq.(102), one gets
\begin{equation}\label{d5}
\Delta^{(0)}(T;\hat{N},\hat{N})=\frac{R^2}{\beta^2}\left[\frac{\beta|m|}{e^{\beta|m|}
+1}+\ln(1+e^{-\beta|m|})\right],
\end{equation}
and, using Eq.(92), one gets \cite{Si4}
%\begin{multline}\label{e17}
\begin{eqnarray}
\Delta^{(1)}(T;\hat N,\hat
N)=\frac{\sin(F\pi)}{2\pi}\int_0^\infty\frac{du}{u}\,{\rm
sech}^2\left(\frac12\beta m\sqrt{u+1}\right)\times \nonumber\\
\times \frac{Fu^FA+(1-F)u^{1-F}A^{-1}-u(2F-1)\cos(F\pi)}
{[u^FA-u^{1-F}A^{-1}+2\cos(F\pi)]^2+4(u+1)\sin^2(F\pi)}+ \nonumber\\
+\frac 1{4}\theta(-\cos \Theta)\,\,{\rm sech}^2\left(\frac12\beta
E_{BS}\right)-\frac14 F(1-F)\,\,{\rm sech}^2\left(\frac12\beta
m\right).
\end{eqnarray}
%\end{multline}

Let us turn now to another conserved observable - angular momentum.
The operator of this observable is given by Eq.(40) and, in this
instance, we take kinetic angular momentum (77) as operator $J$ in
Eq.(40). Using Eq.(103), one gets
\begin{equation}\label{d5}
M^{(0)}(T)=\frac{R^2sm}{\beta}\ln(1+e^{-\beta|m|}),
\end{equation}
and, using Eq.(93), one gets \cite{Si5}
\begin{equation}\label{d5}
M^{(1)}(T)=-s(F-\frac{1}{2})N(T)+\frac{s}{4}F(1-F)\tanh(\frac{1}{2}\beta
m),
\end{equation}
where $N(T)$ is the average fermion number, see Eq.(105). Contrary
to the average angular momentum, the correlation of angular momentum
with fermion number is finite in the infinite volume limit:
\begin{equation}\label{d5}
\Delta(T;\hat{M},\hat{N})=-s(F-\frac{1}{2})\Delta^{(1)}(T;\hat{N},\hat{N})-
\frac{s}{12}(F-\frac{1}{2})F(1-F)\,{\rm sech}^2(\frac{1}{2}\beta m),
\end{equation}
where $\Delta^{(1)}(T;\hat{N},\hat{N})$ is the finite piece of the
fermion number fluctuation, see Eq.(110); note that the latter
relation follows from Eq.(93) and equality
$\Delta^{(0)}(T;\hat{M},\hat{N})=0$ following in its turn from
Eq.(103). The quadratic fluctuation of angular momentum increases as
the volume squared in the large volume limit:
\begin{eqnarray}
\Delta^{(0)}(T;\hat{M},\hat{M})=\frac{R^2}{2\beta^2}\left[
\frac{3R^2}{\beta^2}\int\limits_{\beta|m|}^{\infty}du\,u
\ln(1+e^{-u})+(R^2m^2+\frac{1}{2})\ln(1+e^{-\beta|m|})+\right.
\nonumber
\\ \left.+\frac{1}{2}\frac{\beta|m|}{e^{\beta|m|}+1}\right],
\end{eqnarray}
where Eq.(104) is used. Using Eq.(94), one gets finite piece
\cite{Si5}
\begin{equation}
\Delta^{(1)}(T;\hat{M},\hat{M})=(F-\frac{1}{2})^2\Delta^{(1)}(T;\hat{N},\hat{N})-
\frac{1}{8}F^2(1-F)^2{\rm sech}^2(\frac{1}{2}\beta m).
\end{equation}

Turning to nonconserved observables, see Eq.(52), one can take
$\Upsilon$ to be proportional either to ${\bf x}\cdot {\balpha}$ or
to ${\bf x}\times \balpha$. One can verify that the spectral density
in the first case is identically zero, whereas it is nonzero in the
second case. To be more precise, let us take
\begin{equation}
\Upsilon=\frac{e^2}{4\pi}\,{\bf x}\times \balpha,
\end{equation}
then operator $\hat{O}(52)$ corresponds to the observable with the
physical meaning of the induced magnetic flux multiplied by $e$ and
divided by $2\pi$ (for details see, e.g., Refs.\cite{Si7, Si5}). The
thermal average of this observable is finite in the infinite volume
limit
%\begin{multline}\label{d19}
\begin{eqnarray}
O(T)=-\frac{e^2s F(1-F)}{4\pi m}\left\{\frac{\sin(F\pi)}{\pi}
\int\limits_0^\infty\frac{du}{u\sqrt{u+1}}\tanh\left(\frac12\beta
m\sqrt{u+1}\right)\right.\times \nonumber \\ \times
\frac{u^FA-u^{1-F}A^{-1}-2u\cos(F\pi)}
{[u^FA-u^{1-F}A^{-1}+2\cos(F\pi)]^2+4(u+1)\sin^2(F\pi)}+ \nonumber
\\+\theta(-\cos\Theta)\left.
\frac{m}{(2F\!-\!1)E_{BS}+m}\tanh\left(\frac12\beta
E_{BS}\right)+\frac13\left(F-\frac12\right)\tanh\left(\frac12\beta
m\right)\right\},
\end{eqnarray}
%\end{multline}
as well as its correlation with fermion number
\begin{eqnarray}
\Delta(T;\hat O,\hat N)=\frac{e^2s F(1-F)}{8\pi
m}\left\{\frac{\sin(F\pi)}{\pi}\int_0^\infty\frac{du}{u}\,{\rm
sech}^2\left(\frac12\beta m\sqrt{u+1}\right)\times\right. \nonumber \\
\times \frac{u^FA+u^{1-F}A^{-1}}
{[u^FA-u^{1-F}A^{-1}+2\cos(F\pi)]^2+4(u+1)\sin^2(F\pi)}+ \nonumber \\
\left.+\,\theta(-\cos\Theta)\frac{m}{(2F\!-\!1)E_{BS}+m}\,{\rm
sech}^2\left(\!\frac12\beta E_{BS}\!\right)-\frac{1}{2}\,{\rm
sech}^2\left(\!\frac12\beta m\!\right)\right\}
\end{eqnarray}
and its correlation with angular momentum
\begin{equation}\label{e26}
\Delta(T;\hat O,\hat M)=-s\left(F-\frac12\right) \Delta(T;\hat
O,\hat N),
\end{equation}
see Ref.\cite{Si5}.

Other nonconserved observables are two pieces of angular momentum,
i.e. spin and orbital angular momentum. The ideal gas contribution
to average spin coincides with Eq.(111), thus the average orbital
angular momentum is finite. Expressions for the averages of spin and
orbital angular momentum and their correlations with conserved
observables are given in Ref.\cite{Si5}.

It should be emphasized that the average of induced flux and its
correlations with conserved observables are finite in the infinite
volume limit. The average of fermion number and its correlation with
angular momentum are also finite in this limit, whereas the
fluctuation of fermion number, as well as the average of angular
momentum, diverges as $R^2$, and the fluctuation of angular momentum
diverges as $R^4$. Thus the ideal gas contribution to the thermal
characteristics of angular momentum is predominant, unless
temperature is zero.

In the limit $T\rightarrow0$ ($\beta\rightarrow \infty$) the average
of angular momentum tends to finite value:
\begin{equation}
M(0)=\left\{
\begin{array}{ll}
  \left.\begin{array}{ll}
    \frac{1}{4}s\,{\rm sgn}(m)(1-F)^2,
     & -1<A<\infty  \vspace{0.2cm} \\
    \frac{1}{4}s\,{\rm sgn}(m)F^2, & A^{-1}=0,-1 \vspace{0.2cm} \\
    \frac{1}{4}s\,{\rm sgn}(m)[2F^2-(1-F)^2], & -\infty<A<-1
  \end{array}\right\},
   & 0<F\leq\frac{1}{2}, \vspace{0.2cm} \\
\left.\begin{array}{ll}
    \frac{1}{4}s\,{\rm sgn}(m)F^2,
     & -1<A^{-1}<\infty \vspace{0.2cm} \\
    \frac{1}{4}s\,{\rm sgn}(m)(1-F)^2, & A=0,-1 \vspace{0.2cm} \\
    \frac{1}{4}s\,{\rm sgn}(m)[2(1-F)^2-F^2], & -\infty<A^{-1}<-1
  \end{array}\right\},
   & \frac{1}{2}\leq F<1,
\end{array}\right.
\end{equation}
whereas its fluctuation tends to zero for almost all values of
$\Theta$ with exception of the one corresponding to the zero bound
state energy, $E_{BS}=0$ ($A=-1$):
\begin{equation}
\Delta(0;\hat{M},\hat{M})=\left\{\begin{array}{cl}
                                   0, & A\neq-1, \\
                                   \frac{1}{4}(F-\frac{1}{2})^2, &
                                   A=-1;
                                 \end{array}\right.
\end{equation}
the behaviour of the fermion number fluctuation and the correlations
in this limit is similar to Eq.(121), differing in the values at
$A=-1$. It is instructive to present the average angular momentum in
the form
\begin{equation}
M(T)=M^{(0)}(T)+M(0)+M_{(1)}^{(1)}(T),
\end{equation}
where $M^{(0)}(T)$ and $M(0)$ are given by Eqs.(111) and (120),
respectively, and
\begin{eqnarray}
M_{(1)}^{(1)}(T)=-s(F-\frac{1}{2})\,\theta(-\cos\Theta)\frac{{\rm
sgn}(E_{BS})}{\exp(\beta|E_{BS}|)\!+\!1}-\frac{1}{2}s\left[\frac{1}{2}-F(1\!-\!F)\right]
\!\frac{{\rm sgn}(m)}{\exp(\beta|m|)\!+\!1}- \nonumber \\
\!\!\!\!-s(F\!-\!\frac{1}{2})\frac{\beta
m}{4\pi}\int\limits_{1}^{\infty}\! d w\,{\rm
sech}^2(\frac{1}{2}\beta m w)\arctan\left[
\frac{A(w^2\!-\!1)^F\!-\!A^{-1}(w^2\!-\!1)^{1-F}\!+\!2
\cos(F\pi)}{2w\sin(F\pi)}\right]. \nonumber \\
\end{eqnarray}
Thus, averages, as well as fluctuations and correlations, tend
exponentially to their finite zero-temperature limiting values.

In the high-temperature limit the averages of fermion number and
induced flux tend to zero by power law, see Refs.\cite{Si4, Si5}
respectively, whereas correlations tend to finite values:
\begin{equation}
\Delta(\infty;\hat{M},\hat{N})=\left\{
\begin{array}{lc}
  \left.\begin{array}{lr}
    -\frac{s}{12}(F-\frac{1}{2})(1-F)(3-2F),
     & A^{-1}\neq 0 \vspace{0.1cm}\\
    -\frac{s}{12}(F-\frac{1}{2})F(1+2F), & A^{-1}=0
  \end{array}\right\},
   & 0<F\leq\frac{1}{2}, \vspace{0.1cm} \\
\left.\begin{array}{lr}
    -\frac{s}{12}(F-\frac{1}{2})F(1+2F),
     & \quad A\neq 0 \vspace{0.1cm} \\
    -\frac{s}{12}(F-\frac{1}{2})(1-F)(3-2F), &\quad A=0 \\
      \end{array}\right\},
   & \frac{1}{2}\leq F<1,
\end{array}\right.
\end{equation}
\begin{eqnarray}
\Delta(\infty;\hat{O},\hat{N})= \frac{e^2sF(1-F)}{8\pi m}
\left\{\theta(-\cos \Theta)\frac{m}{(2F-1)E_{BS}+m}-\frac{1}{2}+
 \right.
 \nonumber
\\ \left.+\frac{\sin(F\pi\!)}{\pi}
\int\limits_{0}^{\infty}\!\frac{du}{u}\frac{u^FA+u^{1-F}A^{-1}}
{\left[u^F\!A\!-\!u^{1-F}\!A^{-1}\!+\!2\cos(F\pi)
\right]^2\!+\!4(u\!+\!1)\sin^2(F\pi)}\right\}\!,
\end{eqnarray}
$\Delta(\infty;\hat{O},\hat{M})$ is proportional to Eq.(125), see
Eq.(119). Meanwhile, the average angular momentum, the fermion
number fluctuation, and the angular momentum fluctuation increase as
$T$, $T^2$, and $T^4$, in this limit; such a behaviour is obviously
due to the ideal gas contribution, see Eqs.(111), (109), and (114).

Concluding this section, let us take canonical angular momentum (78)
in the capacity of total angular momentum of the planar system, then
the appropriate operator in the second-quantized theory equals to
$\hat{M}+e\Phi\hat{N}$. It is evident, how to obtain the thermal
characteristics of this observable from our previous results. In
particular, the increasing at large $R$ piece of the average remains
unchanged, Eq.(111), whereas the finite piece of the average takes
form
\begin{equation}
M^{(1)}(T)+e\Phi N(T)=([\![e\Phi]\!]+\frac12)N(T)+\frac
s4F(1-F)\tanh(\frac12\beta m);
\end{equation}
similarly, the correlation with the induced flux takes form
\begin{equation}
\Delta(T;\hat{O},\hat{M}+e\Phi\hat{N})=([\![e\Phi]\!]+\frac12)\Delta(T;\hat{O},\hat{N}).
\end{equation}
The correlation with fermion number consists of two pieces:
\begin{equation}
\Delta^{(0)}(T;\hat{M}+e\Phi\hat{N},\hat{N})=
e\Phi\Delta^{(0)}(T;\hat{N},\hat{N})
\end{equation}
and
\begin{equation}
\Delta^{(1)}(T;\hat{M}+e\Phi\hat{N},\hat{N})=
([\![e\Phi]\!]+\frac12)\Delta^{(1)}(T;\hat{N},\hat{N})-\frac{s}{12}(F-\frac12)
F(1-F){\rm sech}^2(\frac12 \beta m),
\end{equation}
as well as the quadratic fluctuation does:
\begin{equation}
\Delta^{(0)}(T;\hat{M}+e\Phi\hat{N},\hat{M}+e\Phi\hat{N})=
(e\Phi)^2\Delta^{(0)}(T;\hat{N},\hat{N})+\Delta^{(0)}(T;\hat{M},\hat{M})
\end{equation}
and
\begin{eqnarray}
\Delta^{(1)}(T;\hat{M}+e\Phi\hat{N};\hat{M}+e\Phi\hat{N})=
([\![e\Phi]\!]+\frac12)\Delta^{(1)}(T;\hat{N},\hat{N})- \nonumber
\\ -\frac{1}{2}F(1-F)
\left[\frac{s}{3}(F-\frac12)e\Phi+\frac14F(1-F)\right]{\rm
sech}^2(\frac12 \beta m).
\end{eqnarray}
It is straightforward to obtain zero-temperature and
high-temperature limits of the above relations; in particular, the
zero-temperature limit of Eq.(126) was first obtained in
Ref.\cite{Si9}.

\section{Discussion}

In the present paper we consider fractionalization of angular
momentum around a magnetic vortex in the framework of quantum field
theory at finite temperature; for the sake of completeness, the
results for fermion number and induced magnetic flux are also
included. If the kinetic definition of angular momentum is chosen,
then all thermal averages, fluctuations, and correlations are
periodic in the value of the vortex flux, i.e. they depend on the
fractional part of $e\Phi$. If the canonical definition of angular
momentum is chosen, then those thermal characteristics which involve
it depend on both the fractional and integer parts of $e\Phi$. The
difference between two definitions becomes especially striking in
the case of the correlation of angular momentum with fermion number:
the kinetic definition yields finite quantity, see Eq.(113), whereas
the canonical definition yields a piece which increases in the large
volume limit and is proportional to $e\Phi$, see Eq.(128); the
latter looks rather unnatural, since such a piece should correspond
to the ideal gas contribution. The same unnaturalness is relevant
for the fluctuation of the canonically defined angular momentum: its
increasing with the volume size piece which should correspond to the
ideal gas contribution contains a term which is proportional to
$(e\Phi)^2$, see Eq.(130).

According to the generally accepted paradigm, physical
manifestations of the Bohm--Aharonov effect depend exclusively on
the fractional part of $e\Phi$ (see, e.g., Refs.\cite{Pes, SiM,
Aud}), and this favours the kinetic definition of angular momentum.
Moreover, the canonical definition of angular momentum yields rather
embarrassing results for its fluctuation and correlation with
fermion number.

Our analysis has been carried out for the whole variety of boundary
conditions at the location of the vortex, and these conditions are
specified by self-adjoint extension parameter $\Theta$. The
nonvanishing of the angular momentum fluctuation signifies that
angular momentum is not a sharp quantum observable and has to be
understood as a thermal average only. In the high-temperature limit,
angular momentum increases as $T$, see Eq.(111), and its fluctuation
increases as $T^4$, see Eq.(114). In the zero temperature limit,
angular momentum becomes a sharp quantum observable with finite
value $M(0)$ (120). However, the last statement is true for all
values of $\Theta$ in the case of $F=\frac12$, whereas in the case
of $F\neq\frac12$ it is true for almost all values of $\Theta$ with
the exception of one corresponding to the zero bound state energy,
$E_{BS}=0$ ($A=-1$), since in the latter case the zero-temperature
fluctuation is nonzero, see Eq.(121).

Among the whole variety of boundary conditions, let us choose the
condition of minimal irregularity, i.e. the condition corresponding
to the radial components being divergent at $r\rightarrow 0$ at most
as $r^{-p}$ with $p\leq \frac{1}{2}$ \cite{Si0, Si6, Si7}:
\begin{equation}
\Theta=\left\{
  \begin{array}{lc}
    s\frac{\pi}{2}({\rm mod}\,2\pi),
     & 0<F<\frac{1}{2} \vspace{0.2cm} \\
    0({\rm mod}\,2\pi), & F=\frac{1}{2} \vspace{0.2cm} \\
    -s\frac{\pi}{2}({\rm mod}\,2\pi), & \frac{1}{2}<F<1
  \end{array}\right.,
\end{equation}
or $A^{-1}=0$ at $0<F<\frac{1}{2}$, $A=1$ at $F=\frac{1}{2}$, $A=0$
at $\frac{1}{2}<F<1$. Under this condition thermal characteristics
take rather simple form:
\begin{equation}
M^{(1)}(T)=\frac{1}{4}s\left(\frac{1}{2}-|F-\frac{1}{2}|\right)^2
\tanh(\frac{1}{2}\beta m),
\end{equation}
\begin{equation}
\Delta^{(1)}(T;\hat{M},\hat{M})=-\frac{1}{8}\left(\frac{1}{2}-|F-
\frac{1}{2}|\right)^2\left[|F-\frac{1}{2}|+F(1-F)\right]{\rm
sech}^2(\frac{1}{2}\beta m),
\end{equation}
\begin{equation}
\Delta(T;\hat{M},\hat{N})=-\frac{1}{6}s\left(F-
\frac{1}{2}\right)\left(\frac{1}{2}-|F-\frac{1}{2}|\right)
\left(1-|F-\frac{1}{2}|\right){\rm sech}^2(\frac{1}{2}\beta m),
\end{equation}
\begin{equation}
\Delta(T;\hat{O},\hat{M})=0.
\end{equation}

We recall that the results of the present paper are relevant for
planar fermions in an irreducible $2\times 2$ representation of the
Clifford algebra in $2+1$-dimensional space-time, see Eqs.(75) and
(76); the mass of such fermions violates parity. As it should be
expected, our results remain invariant under transitions to
equivalent representations of the Clifford algebra, i.e. are
independent of $\chi_s$. As to a transition to the inequivalent
representation ($s\rightarrow -s$ or $m\rightarrow -m$), the
averages of fermion number and angular momentum are odd and the
average induced flux is even under this transition, see Eqs.(105),
(111), (112), and (117); the quadratic fluctuations of fermion
number and angular momentum, as well as their correlation, are even
and the correlations of induced flux with fermion number and angular
momentum are odd under this transition, see Eqs.(109), (110),
(113)-(115), (118), and (119).

Finally, let us discuss consequences for planar fermions with
parity-conserving mass; such fermions are assigned to a reducible
$4\times 4$ representation which is composed as a direct sum of two
inequivalent irreducible representations. Thermal characteristics in
the reducible representation are obtained by summing those in the
irreducible one over $s=\pm 1$. In particular, the average induced
flux takes form
\begin{equation}
O_{4\times 4}(T)=2\,O(T)\biggl|_{\begin{array}{l}
                                           \scriptstyle F=e\Phi-[\![e\Phi]\!] \\
                                           [-0.2cm]\scriptstyle s=1 \\ [-0.2cm]
                                           \scriptstyle m=|m| \\
                                        \end{array}},
\end{equation}
where $O(T)$ is given by Eq.(117). The averages of fermion number
and angular momentum, as well as their correlations with induced
flux, are vanishing,
\begin{equation}
N_{4\times 4}(T)=M_{4\times 4}(T)=0,\,\,\,\, \Delta_{4\times
4}(T;\,\hat{O},\hat{N})=\Delta_{4\times 4}(T;\,\hat{O},\hat{M})=0,
\end{equation}
while the fluctuations of fermion number and angular momentum, as
well as their correlation, are nonvanishing,
\begin{equation}
\Delta_{4\times 4}(T;\,\hat{N},\hat{N})=
2\,\Delta(T;\,\hat{N},\hat{N})\biggl|_{\begin{array}{l}
                                           \scriptstyle F=e\Phi-[\![e\Phi]\!] \\
                                           [-0.2cm]\scriptstyle s=1 \\ [-0.2cm]
                                           \scriptstyle m=|m| \\
                                        \end{array}},
\end{equation}
\begin{equation}
\Delta_{4\times 4}(T;\,\hat{M},\hat{M})=
2\,\Delta(T;\,\hat{M},\hat{M})\biggl|_{\begin{array}{l}
                                           \scriptstyle F=e\Phi-[\![e\Phi]\!] \\
                                           [-0.2cm]\scriptstyle s=1 \\ [-0.2cm]
                                           \scriptstyle m=|m| \\
                                        \end{array}},
\end{equation}
\begin{equation}
\Delta_{4\times 4}(T;\,\hat{M},\hat{N})=
2\,\Delta(T;\,\hat{M},\hat{N})\biggl|_{\begin{array}{l}
                                           \scriptstyle F=e\Phi-[\![e\Phi]\!] \\
                                           [-0.2cm]\scriptstyle s=1 \\ [-0.2cm]
                                           \scriptstyle m=|m| \\
                                        \end{array}}.
\end{equation}
Note that Eqs.(137) and (141) are finite in the large volume limit,
thus depending essentially on the fractional part of $e\Phi$,
whereas the dependence of Eqs.(139) and (140) on the vortex flux is
not essential, since the ideal gas contribution to fluctuations is
nonzero and prevailing in this limit. Obviously, our arguments in
favour of the kinetic definition of angular momentum remain valid in
this case also.

\renewcommand{\thesection}{}
\renewcommand{\theequation}{\thesection.\arabic{equation}}
\setcounter{section}{1} \setcounter{equation}{0}

\section*{Acknowledgements}

This research was supported in part by the Swiss National Science
Foundation under the SCOPES project IB7320-110848. N.D.V. was
supported by the INTAS Fellowship Grant for Young Scientists.

\renewcommand{\thesection}{A}
\renewcommand{\theequation}{\thesection.\arabic{equation}}
\setcounter{section}{1} \setcounter{equation}{0}

\section*{Appendix A. Canonical and improved definitions of the moments
of momentum tensor of classical fields}

Let $\widetilde{T}^{\mu\nu}$ be the canonical tensor of energy and
momentum of classical fields, and let us define tensor
\begin{equation}
T^{\mu\nu}=\widetilde{T}^{\mu\nu}+\partial_\lambda\chi^{\mu\nu\lambda},
\end{equation}
where $\chi^{\mu\nu\lambda}$ is an arbitrary third-rank tensor
which is antisymmetric in two last indices,
$\chi^{\mu\nu\lambda}=-\chi^{\mu\lambda\nu}$. Then
\begin{equation}
\partial_\nu T^{\mu\nu}=\partial_\nu \widetilde{T}^{\mu\nu},
\end{equation}
and conservation of $\widetilde{T}^{\mu\nu}$ leads to that of
$T^{\mu\nu}$. Moreover, both tensors yield the same
energy-momentum vector
\begin{equation}
P^\mu=\int d^3x\,\widetilde{T}^{\mu 0} =\int d^3x\, T^{\mu 0},
\end{equation}
since the contribution of $\chi^{\mu\nu\lambda}$ is transformed
into an integral over a surface enclosing space,
\begin{equation}
\int d^3x\, \partial_\lambda \chi^{\mu 0\lambda}=\int d^3x\,
\partial_l \chi^{\mu 0l}=\oint d\sigma^l\chi^{\mu 0l},
\end{equation}
and the latter integral vanishes under mild assumptions on the
decrease of $\chi^{\mu 0l}$ at large distances.

Similarly, taking the canonical tensor of densities of moments of
momentum
\begin{equation}
\widetilde{M}^{\mu \nu;\,\rho}=x^{\mu}\widetilde{T}^{\nu \rho}
-x^{\nu}\widetilde{T}^{\mu \rho}+S^{\mu \nu;\,\rho},
\end{equation}
one defines tensor
\begin{equation}
M^{\mu\nu;\,\rho}=\widetilde{M}^{\mu\nu;\,\rho}+\partial_\lambda(x^{\mu}
\chi^{\nu\rho\lambda}-x^{\nu}\chi^{\mu\rho\lambda}),
\end{equation}
which satisfies divergence relation
\begin{equation}
\partial_\rho M^{\mu\nu;\,\rho}=\partial_\rho
\widetilde{M}^{\mu\nu;\,\rho},
\end{equation}
and conservation of $\widetilde{M}^{\mu\nu;\,\rho}$ leads to that
of $M^{\mu\nu;\,\rho}$.

In cases when tensors $\widetilde{T}^{\mu\nu}$ and
$\widetilde{M}^{\mu\nu;\,\rho}$ fail to meet some plausible
requirements, the transition to tensors $T^{\mu\nu}$ and
$M^{\mu\nu;\,\rho}$ may allow one to eliminate such failings. In
particular, the canonical energy-momentum tensor of the
electromagnetic field is neither symmetric nor gauge invariant. By
choosing a concrete form of $\chi^{\mu\nu\lambda}$ one can remove
both of these shortcomings. Incidentally, a convenient expression
which does not involve spin part $S^{\mu\nu;\,\rho}$ is obtained for
the moments of momentum tensor (Belinfante's theorem \cite{Bel}).

Namely, first, Eq.(A.6) with the use of Eqs.(A.1) and (A.5) is
recast into the form
\begin{equation}
M^{\mu\nu;\,\rho}=x^\mu T^{\nu\rho}-x^\nu
T^{\mu\rho}+\chi^{\nu\rho\mu}-\chi^{\mu\rho\nu}+S^{\mu\nu;\,\rho}.
\end{equation}
Then defining $\chi$-tensor as
\begin{equation}
\chi^{\mu\nu\lambda}=\frac{1}{2}(-S^{\mu\nu;\,\lambda}+
S^{\nu\lambda;\,\mu}-S^{\lambda\mu;\,\nu})
\end{equation}
(recall that $S$-tensor is antisymmetric in the first two indices
and $\chi$-tensor is antisymmetric in the last two indices), one
gets
\begin{equation}
M^{\mu\nu;\,\rho}=x^\mu T^{\nu\rho}-x^\nu T^{\mu\rho}.
\end{equation}
Thus one gets following expressions for the canonical tensor of
moments of momentum
\begin{equation}
\widetilde{M}^{\mu\nu}=\int d^3x\,[x^\mu T^{\nu 0}-x^\nu T^{\mu
0}-\partial_\lambda(x^\mu \chi^{\nu 0\lambda}-x^\nu \chi^{\mu
0\lambda})]
\end{equation}
and the improved one
\begin{equation}
M^{\mu\nu}=\int d^3x\,(x^\mu T^{\nu 0}-x^\nu T^{\mu 0}).
\end{equation}
Although the difference between Eqs.(A.12) and (A.11) is
transformed as previously, see Eq.(A.4), into an integral over a
closed surface,
\begin{equation}
\int d^3 x\,\partial_\lambda(x^\mu\chi^{\nu
0\lambda}-x^\nu\chi^{\mu 0\lambda})=\oint d\sigma^l(x^\mu
\chi^{\nu 0 l}-x^\nu \chi^{\mu 0 l}),
\end{equation}
this integral may appear to be finite for certain, long-range,
field configurations, since it contains an additional power of
large distance, $|\bf{x}|$. In this case one has to decide,
whether Eq.(A.11) or Eq.(A.12) gives physically meaningful moments
of momentum.

Let us consider a system characterized by lagrangian
\begin{equation}
{\cal{L}}=-\frac{1}{4}F^{\mu\nu}F_{\mu\nu}-A^\mu j_{\mu},
\end{equation}
where $F_{\mu\nu}=\partial_\mu A_\nu -\partial_\nu A_\mu$ is the
electromagnetic field strength, $A_\mu$ is the appropriate vector
potential, $j_\mu$ is the charge current. The improved
energy-momentum tensor of the system is proved to be \cite{Bel} (see
also, e.g., Ref.\cite{Itzyk})
\begin{equation}
T^{\mu\nu}=\frac{1}{4}g^{\mu\nu}F^{\lambda\rho}F_{\lambda\rho}+
F^{\mu\lambda}{F_{\lambda}}^{\nu}+g^{\mu\nu}A^\lambda
j_\lambda-A^\mu j^\nu
\end{equation}
(note that purely electromagnetic part is symmetric and gauge
invariant); incidentally, one gets
\begin{equation}
\chi^{\mu\nu\lambda}=A^\mu F^{\nu\lambda}.
\end{equation}
Defining an axial vector from spatial tensor components
\begin{equation}
M^i=\frac{1}{2}\varepsilon^{ikl}M_{kl},
\end{equation}
and using Eq.(A.12), one gets the following expression for the
improved angular momentum
\begin{equation}
{\bf{M}}=\int d^3x\,[({\bf{x}}\times({\bf{E}}\times
{\bf{B}}))-({\bf{x}}\times {\bf{A}}) j^0],
\end{equation}
where $E^i=F^{i0}$ and $B^i=-\frac{1}{2}\varepsilon^{ikl}F_{kl}$
are the electric and magnetic field strengths.

\renewcommand{\thesection}{B}
\renewcommand{\theequation}{\thesection.\arabic{equation}}
\setcounter{section}{1} \setcounter{equation}{0}

\section*{Appendix B. Derivation of representation (45) for
thermodynamic potential}

The partition function of the fermionic system is presented as the
functional integral over the Grassman fields
\begin{equation}
\exp[-\beta\Omega(\beta,\mu)]=\int d\psi^+d\psi \,e^{-S},
\end{equation}
where
\begin{equation}
S=\int\limits_{0}^{\beta}d\tau\int d^d x\,\psi^+(\partial_\tau- \mu
J+H)\psi
\end{equation}
is the Euclidean action, $\tau$ is the imaginary time. The integral
in Eq.(B.1) is of the Gauss type and can be immediately computed
\begin{equation}
\exp[-\beta\Omega(\beta,\mu)]={\rm det}(\partial_\tau-\mu J+H).
\end{equation}
Hence the thermodynamic potential is given by expression
$$
\Omega(\beta,\mu)=-\frac{1}{\beta}\ln {\rm det}(\partial_\tau -\mu
J+H)=
$$
\begin{equation} =-\frac{1}{\beta}\int\limits_{0}^{\beta} d\tau\int d^dx
\,{\rm tr}\langle{\bf x},\tau|\ln(\partial_\tau-\mu J+H)|{\bf
x,\tau}\rangle.
\end{equation}
In the case of a static background, operators $H$ and $J$ are
$\tau$-independent, and the integration over $\tau$ is performed by
using the antiperiodicity boundary condition at the ends of the
imaginary time interval:
\begin{equation}
\Omega(\beta,\mu)=-\frac{1}{\beta}\sum\limits_{n=-\infty}^{\infty}
\int d^d x\,{\rm tr}\langle{\bf x}|\ln(H-\mu J-i\omega_n)|{\bf
x}\rangle,
\end{equation}
where $\omega_n=\frac{2\pi}{\beta}(n+\frac{1}{2})$, and summation is
over integer values of $n$. Using the notation of functional trace,
one gets further
$$
\Omega(\beta,\mu)=-\frac{1}{\beta}\sum\limits_{n=-\infty}^{\infty}
{\rm Tr}\ln(H-\mu J-i\omega_n)=-\frac{1}{\beta}{\rm Tr}\ln \prod
\limits_{n=0}^{\infty}[(H-\mu J)^2+\omega_n^2]=
$$
\begin{equation}
=-\frac{1}{\beta}\left\{{\rm
Tr}\ln\prod\limits_{n=0}^{\infty}\left[\left( \frac{H-\mu
J}{\omega_n}\right)^2+1\right]+{\rm
Tr}\ln\prod\limits_{n=0}^{\infty} \omega_n^2\right\}.
\end{equation}
The second term in the figure brackets is dropped as an irrelevant
infinite constant, and the infinite product in the first term is
computed with the use of relation \cite{Prud}
$$
{\rm cosh}(\frac{\pi
a}{2})=\prod\limits_{n=0}^{\infty}\left[1+a^2(2n+1)^{-2}\right].
$$
As a result we get expression (45) for the thermodynamic potential.

\renewcommand{\thesection}{C}
\renewcommand{\theequation}{\thesection.\arabic{equation}}
\setcounter{section}{1} \setcounter{equation}{0}

\section*{Appendix C. Radial components of $G^\omega(r,\varphi;\,r',\varphi')$}

The radial components of the resolvent kernel of $H$ (73) take form
(see Ref.\cite{Si4}),

type 1 $(l=s(n-n_c)>0)$:
\begin{equation}\label{ap1}
a_n(r;r')=\frac{i\pi}2(\omega+m)\left[\theta(r-r')H^{(1)}_{l-F}(k
r)J_{l-F}(k r')+ \theta(r'-r)J_{l-F}(k r)H^{(1)}_{l-F}(k r')\right],
\end{equation}
\begin{equation}\label{ap2}
b_n(r;r')=\frac{i\pi}2k\left[\theta(r-r')H^{(1)}_{l+1-F}(k
r)J_{l-F}(k r')+ \theta(r'-r)J_{l+1-F}(k r)H^{(1)}_{l-F}(k
r')\right],
\end{equation}
\begin{equation}\label{ap3}
c_n(r;r')=\frac{i\pi}2(\omega-m)\left[\theta(r-r')H^{(1)}_{l+1-F}(k
r)J_{l+1-F}(k r')+ \theta(r'-r)J_{l+1-F}(k r)H^{(1)}_{l+1-F}(k
r')\right],
\end{equation}
\begin{equation}\label{ap4}
d_n(r;r')=\frac{i\pi}2k\left[\theta(r-r')H^{(1)}_{l-F}(k
r)J_{l+1-F}(k r')+ \theta(r'-r)J_{l-F}(k r)H^{(1)}_{l+1-F}(k
r')\right];
\end{equation}

type 2 $(l'=-s(n-n_c)>0)$:
\begin{equation}\label{ap5}
a_n(r;r')=\frac{i\pi}2(\omega+m)\left[\theta(r-r')H^{(1)}_{l'+F}(k
r)J_{l'+F}(k r')+ \theta(r'-r)J_{l'+F}(k r)H^{(1)}_{l'+F}(k
r')\right],
\end{equation}
\begin{equation}\label{ap6}
b_n(r;r')=-\frac{i\pi}2k\left[\theta(r-r')H^{(1)}_{l'-1+F}(k
r)J_{l'+F}(k r')+ \theta(r'-r)J_{l'-1+F}(k r)H^{(1)}_{l'+F}(k
r')\right],
\end{equation}
\begin{equation}\label{ap7}
c_n(r;r')=\frac{i\pi}2(\omega-m)\left[\theta(r-r')H^{(1)}_{l'-1+F}(k
r)J_{l'-1+F}(k r')+ \theta(r'-r)J_{l'-1+F}(k r)H^{(1)}_{l'-1+F}(k
r')\right],
\end{equation}
\begin{equation}\label{ap8}
d_n(r;r')=-\frac{i\pi}2k\left[\theta(r-r')H^{(1)}_{l'+F}(k
r)J_{l'-1+F}(k r')+ \theta(r'-r)J_{l'+F}(k r)H^{(1)}_{l'-1+F}(k
r')\right];
\end{equation}

type 3 $(n=n_c)$:
\newline
$$
a_{n_c}(r;r')=\frac{i\pi}2\frac{\omega+m}{\sin\nu_\omega+\cos\nu_\omega
e^{iF\pi}}\left\{\theta(r-r')H^{(1)}_{-F}(k r) [\sin\nu_\omega
J_{-F}(k r')+\cos\nu_\omega J_{F}(k r')]+\right.
$$
\begin{equation}
\left.+\theta(r'-r)[\sin\nu_\omega J_{-F}(k r)+\cos\nu_\omega
J_{F}(k r)]H^{(1)}_{-F}(k r') \right\},
\end{equation}
%\end{multline}
%\begin{multline}\label{ap10}
$$
b_{n_c}(r;r')=\frac{i\pi}2\frac{k}{\sin\nu_\omega+\cos\nu_\omega
e^{iF\pi}}\left\{\theta(r-r')H^{(1)}_{1-F}(k r) [\sin\nu_\omega
J_{-F}(k r')+\cos\nu_\omega J_{F}(k r')]+\right.
$$
\begin{equation}
+\left.\theta(r'-r)[\sin\nu_\omega J_{1-F}(k r)-\cos\nu_\omega
J_{-1+F}(k r)]H^{(1)}_{-F}(k r') \right\},
\end{equation}
%\end{multline}
%\begin{multline}\label{ap11}
$$
c_{n_c}(r;r')=\frac{i\pi}2\frac{\omega-m}{\sin\nu_\omega+\cos\nu_\omega
e^{iF\pi}}\left\{\theta(r-r')H^{(1)}_{1-F}(k r) [\sin\nu_\omega
J_{1-F}(k r')-\cos\nu_\omega J_{-1+F}(k r')]+\right.
$$
\begin{equation}
+\left.\theta(r'-r)[\sin\nu_\omega J_{1-F}(k r)-\cos\nu_\omega
J_{-1+F}(k r)]H^{(1)}_{1-F}(k r') \right\}\,,
\end{equation}
%\end{multline}
%\begin{multline}\label{ap12}
$$
d_{n_c}(r;r')=\frac{i\pi}2\frac{k}{\sin\nu_\omega+\cos\nu_\omega
e^{iF\pi}}\left\{\theta(r-r')H^{(1)}_{-F}(k r) [\sin\nu_\omega
J_{1-F}(k r')-\cos\nu_\omega J_{-1+F}(k r')]+\right.
$$
\begin{equation}
+\left.\theta(r'-r)[\sin\nu_\omega J_{-F}(k r)+\cos\nu_\omega
J_{F}(k r)]H^{(1)}_{1-F}(k r') \right\}.
\end{equation}
%\end{multline}
Here $J_\rho(u)$ is the Bessel function of order $\rho$,
$H^{(1)}_\rho(u)$ is the first-kind Hankel function of order $\rho$,
and
\begin{equation}\label{ap13}
\tan\nu_\omega=\frac{k^{2F}}{\omega+m}\,{\rm{sgn}}(m)(2|m|)^{1-2F}
\frac{\Gamma(1-F)}{\Gamma(F)}\tan\left(s\frac\Theta2+\frac\pi4\right)\,.
\end{equation}

In the absence of the vortex the radial components take form:
\begin{equation}\label{ap14}
\left.a_n(r;r')\right|_{e\Phi=0}=\frac{i\pi}2(\omega+m)\left[\theta(r-r')H^{(1)}_{sn}(k
r) J_{sn}(k r')+\theta(r'-r)J_{sn}(k r)H^{(1)}_{sn}(k r')\right],
\end{equation}
\begin{equation}\label{ap15}
\left.b_n(r;r')\right|_{e\Phi=0}=\frac{i\pi}2k\left[\theta(r-r')H^{(1)}_{sn+1}(k
r) J_{sn}(k r')+\theta(r'-r)J_{sn+1}(k r)H^{(1)}_{sn}(k r')\right],
\end{equation}
\begin{equation}\label{ap16}
\left.c_n(r;r')\right|_{e\Phi=0}=\frac{i\pi}2(\omega-m)\left[\theta(r-r')H^{(1)}_{sn+1}(k
r) J_{sn+1}(k r')+\theta(r'-r)J_{sn+1}(k r)H^{(1)}_{sn+1}(k
r')\right],
\end{equation}
\begin{equation}\label{ap17}
\left.d_n(r;r')\right|_{e\Phi=0}=\frac{i\pi}2k\left[\theta(r-r')H^{(1)}_{sn}(k
r) J_{sn+1}(k r')+\theta(r'-r)J_{sn}(k r)H^{(1)}_{sn+1}(k
r')\right].
\end{equation}

\renewcommand{\thesection}{D}
\renewcommand{\theequation}{\thesection.\arabic{equation}}
\setcounter{section}{1} \setcounter{equation}{0}

\section*{Appendix D. Derivation of Eqs.(96)-(98)}

Using integral representation of kernel (95)
\begin{equation}
\langle{\bf
x}\left|\left(H_0-\omega\right)^{-1}e^{-tH_0^2}\right|{\bf
x}'\rangle=\int\frac{d^2p}{(2\pi)^2}\exp\left[i{\bf p}\cdot({\bf
x}-{\bf x}')-t(p^2+m^2)\right]\frac{{\balpha}\cdot{\bf
p}+\gamma^0m+\omega}{p^2-k^2},
\end{equation}
we calculate its trace over spinor indices at ${\bf x}'={\bf x}$
\begin{equation}
tr\left\langle{\bf x}\left|(H_0-\omega)^{-1}e^{-tH_0^2}\right|{\bf
x}\right\rangle=2\omega\int\frac{d^2p}{(2\pi)^2}\frac{e^{-t(p^2+m^2)}}{p^2-k^2}.
\end{equation}
Similarly, using relations
$$
J_0\left\langle{\bf x}\left|(H_0-\omega)^{-1}e{-tH_0^2}\right|{\bf
x}'\right\rangle=\int\frac{d^2p}{(2\pi)^2}\exp\left[i{\bf
p}\cdot({\bf x}-{\bf x}')-t(p^2+m^2)\right]\times
$$
\begin{equation}
\times\left(x^1p^2-x^2p^1+\frac{1}{2}s\gamma^0\right)\frac{{\balpha}\cdot{\bf
p}+\gamma^0m+\omega}{p^2-k^2},
\end{equation}
and
$$
J_0^2\left\langle{\bf
x}\left|(H_0-\omega)^{-1}e^{-tH_0^2}\right|{\bf
x}'\right\rangle=\int\frac{d^2p}{(2\pi)^2}\exp\left[i{\bf
p}\cdot({\bf x}-{\bf x}')-t(p^2+m^2)\right]\times
$$
\begin{equation}
\times\left(x^1p^2-x^2p^1+\frac{1}{2}s\gamma^0\right)^2\frac{{\balpha}\cdot
{\bf p}+\gamma^0m+\omega}{p^2-k^2},
\end{equation}
we find relations
\begin{equation}
tr\,J_0\left\langle{\bf
x}\left|(H_0-\omega)^{-1}e^{-tH_0^2}\right|{\bf
x}\right\rangle=sm\int\frac{d^2p}{(2\pi)^2}\frac{e^{-t(p^2+m^2)}}{p^2-k^2},
\end{equation}
and
\begin{equation}
tr\,J_0^2\left\langle{\bf
x}\left|(H_0-\omega)^{-1}e^{-tH_0^2}\right|{\bf
x}\right\rangle=\omega\int\frac{d^2p}{(2\pi)^2}(r^2p^2+
\frac{1}{2})\frac{e^{-t(p^2+m^2)}}{p^2-k^2}.
\end{equation}
Eqs.(D.2), (D.5), and (D.6) in the case of $Im\,k>|{\rm Re}\,k|$ are
reduced to Eqs.(96)-(98).

%\newpage

\end{document}